\documentclass[10pt,reprint,tightenlines,onecolumn,amssymb,nobibnotes, aps, prd]{revtex4-2}
\usepackage[paperwidth=210mm,paperheight=297mm,centering,hmargin=2cm,vmargin=2.5cm]{geometry}
\pdfoutput=1
\setcitestyle{authoryear,round}
\usepackage[english]{babel}
\usepackage{graphicx}
\usepackage{xcolor}

\usepackage{amsmath}
\usepackage{xfrac}
\newcommand{\dbar}{d\hspace*{-0.08em}\bar{}\hspace*{0.1em}}
\usepackage{bm}
\usepackage{hyperref}
\hypersetup{
    bookmarks=true,         
    unicode=false,          
    pdftoolbar=true,        
    pdfmenubar=true,        
    pdffitwindow=false,     
    pdfstartview={FitH},    
    pdftitle={Evolution TANN and the discovery of the internal variables and evolution equations in solid mechanics)},    
    pdfauthor={F. Masi, I. Stefanou},     
    pdfnewwindow=true,      
    colorlinks=true,       
    linkcolor=blue,          
    citecolor=blue,        
    filecolor=blue,         
    urlcolor=blue        
}
\usepackage[font=small,labelfont=bf]{caption}
\usepackage{subcaption}
\usepackage{caption}
\begin{document}
\title{\Large \textbf{Evolution TANN and the identification of internal variables and evolution equations in solid mechanics}\footnote{Post-print version of: \textcolor{blue}{F Masi, I Stefanou. “Evolution TANN and the identification of internal variables and evolution equations in solid mechanics”. In: \textit{J Mech Phys Solids} (2023), 105245. doi: \href{https://doi.org/10.1016/j.jmps.2023.105245}{10.1016/j.jmps.2023.105245}.}}}

\author{Filippo Masi\footnote{Correspondence to \textit{Sydney Centre in Geomechanics and Mining Materials, School of Civil Engineering, The University of Sydney, 2006, Sydney, Australia.}}}
 \email{filippo.masi@sydney.edu.au}
\author{Ioannis Stefanou}
 \email{ioannis.stefanou@ec-nantes.fr}
\affiliation{Nantes Université, Ecole Centrale Nantes, CNRS, Institut de Recherche en Génie Civil et Mécanique (GeM), UMR 6183, F-44000, Nantes, France.}%

\date{\today}

\begin{abstract}
Data-driven and deep learning approaches have demonstrated to have the potential of replacing classical constitutive models for complex materials, displaying path-dependency and possessing multiple inherent scales. Yet, the necessity of structuring constitutive models with an incremental formulation has given rise to data-driven approaches where physical quantities, e.g. deformation, blend with artificial, non-physical ones, such as the increments in deformation and time. Neural networks and the consequent constitutive models depend, thus, on the particular incremental formulation, fail in identifying material representations locally in time, and suffer from poor generalization.

Herein, we propose a new approach which allows, for the first time, to decouple the material representation from the incremental formulation. Inspired by the Thermodynamics-based Artificial Neural Networks (TANN) and the theory of the internal variables, the evolution TANN (\textit{e}TANN) are continuous-time and, therefore, independent of the aforementioned artificial quantities. Key feature of the proposed approach is the identification of the evolution equations of the internal variables in the form of ordinary differential equations, rather than in an incremental discrete-time form.\\
In this work, we focus attention to juxtapose and show how the various general notions of solid mechanics are implemented in \textit{e}TANN. The laws of thermodynamics are hardwired in the structure of the network and allow predictions which are always consistent, independently of the range of the training dataset.\\
Inspired by previous works, we propose a methodology that allows to identify, from data and first principles, admissible sets of internal variables from the microscopic fields in complex materials.\\
The capabilities as well as the scalability of the proposed approach are demonstrated through several applications involving a broad spectrum of complex material behaviors, from plasticity to damage and viscosity (and combination of them). Finally, we show that the proposed approach can be used to speed-up state-of-the-art multiscale analyses, by virtue of asymptotic homogenization. \textit{e}TANN provide excellent results compared to detailed fine-scale simulations and offer the possibility not only to describe the average macroscopic material behavior, but also micromechanical, complex mechanisms.
\end{abstract}

\maketitle

\section{Introduction}

\noindent Accurate models for the behavior of materials are of fundamental importance in material science and mechanics. They allow for understanding of mechanical processes and bring the potential of investigating intricate systems by means of numerical tools.

Heuristic constitutive models, traditionally derived from first principles (thermodynamics) and empirical approaches (to ensure calibration over experiments), can hardly describe the behavior of complex materials that display path-dependency and possess multiple inherent scales, e.g. metamaterials.

However, data-driven and machine learning approaches have recently demonstrated to be able to overcome some of the issues encountered in heuristic constitutive modeling. A large amount of data, available from fine-scale simulations and/or experiments, as well as the success of deep learning, have been the impetus for the development of neural networks to identify material constitutive representations and surrogate models \citep[see][for a comprehensive review]{liu2021review}.\\
The high interpolation capabilities of neural networks \citep[NN, see e.g.][]{cybenko1989approximation,chen1995universal} enable the modeling of several aspects of the behavior of complex materials and their microstructure(s), both in elasticity \citep[see e.g.][]{linka2021constitutive} and in inelasticity \citep[see][among others]{mianroodi2021teaching,liu2021review,jones2022neural}. In parallel, neural networks enable massive computational accelerations in otherwise demanding numerical analyses \citep{Feyel2003,Nguyen2014}, i.e., multiscale simulations \citep{le2015computational,liu2019deep,ghavamian2019accelerating,lu2019data,peng2021multiscale,masistefanou2021}.\\

Yet, constitutive modeling based on neural networks faces two major issues: (in-)consistency and (poor) generalization. Consistency refers here to the fulfillment of first principles, derived from physics. Indeed, independently of the specific approach, the identified material representations and constitutive models must obey the laws of thermodynamics \citep{bower2009applied}. Generalization refers, instead, to the ability of predicting material responses in conditions and/or under loading that do not belong to the dataset used for training neural networks. The two issues limit the applicability of classical \textit{black-box} (i.e., physics-agnostic) approaches \citep[cf.][]{karniadakis2021physics,masi2021thermodynamics,liu2021review,cueto2022thermodynamics}. \\

When dealing with constitutive modeling and neural networks, the choice of an adequate state space of the material--that is, the set of variables its constitutive behavior depends on--is of paramount importance. Two different approaches have been widely adopted. The first one, identified under the name of rational thermodynamics \citep{truesdellrational}, assumes, relying on the axiom of fading memory \citep{coleman1961foundations}, that the material behavior at time $t$ depends on the past history of the state variables, i.e., temperature and deformation.\\
The second approach, the so-called theory (or thermodynamics) of internal state variables \citep{coleman1967thermodynamics,maugin1994thermodynamicsA}, has, instead, the advantage of delivering a material description local in time. It assumes that the state of a material at time $t$ can be described by the state variables and an additional set of internal (state) variables, at time $t$. The latter are introduced to describe irreversible and complex mechanisms which are not captured by the state variables, e.g. plasticity, damage, and other micromechanical mechanisms. The identification of the evolution in time of the internal variables, the so-called evolution equations, allows to describe the material response in time. \\

In structuring neural networks-based constitutive models, both ways of defining the state space can be employed. The first case, where the actual state depends on process histories, finds straightforward applications in recurrent neural networks. These models are generally trained to predict the material state at the present time, from a temporal sequence of the state variables \citep[see][among others]{ghavamian2019accelerating,mozaffar2019deep,HEIDER2020112875,bonatti2021one,wu2022recurrent}. However, there is no mathematical proof that \textit{black-box} recurrent and feed-forward neural networks \citep[see][among others]{ghaboussi1998new,lefik2003artificial,jung2006neural,yun2008new,lefik2009artificial} can succeed in the prediction of future events that are not represented in the training set and respect the laws of physics.

Recent developments have risen to tackle the lack of physical consistency in neural networks and the consequent constitutive models. In particular, it has been demonstrated that is possible to structure neural networks in a way that they deliver, by construction, thermodynamics-/physics-consistent predictions. This is the case of the so-called physics- and/or thermodynamics-informed neural networks \citep[see, among others,][]{karniadakis2021physics,hernandez2021deep,masi2021thermodynamics,rocha2021deepbnd,klein2022polyconvex,yin2022interfacing,cueto2022thermodynamics}. Among these approaches, the Thermodynamics-based Artificial Neural Networks \citep[TANN,][]{masi2021thermodynamics,masistefanou2021}, based on the theory of internal variables, are able to uncover constitutive equations from the laws of thermodynamics. 
In parallel, TANN and the aforementioned physics-/thermodynamics- informed neural networks show good generalization capabilities with respect to \textit{black-box} neural networks. This is a direct consequence of the physically-consistent architecture which enables accurate predictions of the material behavior even for unseen loading paths, both within the range of the training set (interpolation) and outside (extrapolation). For more, we refer to \citet{masi2021thermodynamics,masistefanou2021}.

Despite the capabilities of such neural networks, the consequent constitutive models do not necessarily possess the same degree of generalization. The reason lies in the necessity of formulating constitutive models in an incremental formulation, allowing a straightforward implementation in classical Finite Element (FE) codes. As a result, neural networks will also be based on an incremental formulation \citep[i.e., discrete-time,][]{strogatz2018nonlinear}, bringing a formulation of the material behavior where strain, strain increments, and time steps blend together \citep{ghaboussi1998new,lefik2003artificial,jung2006neural,yun2008new,lefik2009artificial,masi2021thermodynamics}. The predictions of the neural network will thus depend on both physical quantities, e.g. strain, and artificial ones, such as strain increments and, in the case of rate-dependent materials, the time step. In this context, it is clear that a high degree of generalization can only be obtained with training datasets that adequately span not only the (real) state space of the material, but also the artificial one, e.g., the range of strain increments.\\

The aim of this work is to extend the methodology of TANN by enabling the decoupling of the material representation and the incremental formulation. We rely on the previous developments and the philosophy of TANN and propose a new version where the representation of the material behavior is local in time. The new approach is referred to as evolution TANN (\textit{e}TANN) to emphasize the fact that the material model allows to identify the evolution equations of the internal variables, rather than the link between the increments of the internal variables and the strain increments. \textit{e}TANN are continuous-time and, therefore, independent of the artificial quantities corresponding to the incremental formulation. The description of the material response is achieved by the numerical time integration of the evolution equations, which is performed outside of the network with well-established techniques.

Two configurations of \textit{e}TANN are proposed, depending on whether the internal variables are priori selected or not (unknown). In the former, the material description passes from the identification of the energy density and of the evolution equations (ordinary differential equations). In the latter, the procedure proposed in \citet{masistefanou2021} is used and extended to avoid a discrete-time dynamics description. Accordingly, thermodynamically consistent internal variables and their corresponding evolution equations are automatically identified from the knowledge of those variables describing the material behavior at the microscopic scale, i.e., the internal coordinates (e.g., displacement, velocity, microscopic deformation fields). In addition, the proposed approach makes possible to identify not only the evolution equations of the internal variables, but also those of the internal coordinates. As a result, \textit{e}TANN are found to describe the evolution of both the average material response and the microscopic one.\\
Finally, the proposed approach results in a reliable and scalable tool to speed-up otherwise computationally prohibitive fine-scale simulations, relying on a multiscale approach and scales separation \citep[see][]{sanchez1986homogenization,Bakhvalov1989,pinho2009asymptotic}.\\

The paper is structured as follows. \hyperref[sec:thermodynamics]{Section~\ref*{sec:thermodynamics}} reviews the theoretical foundations (neural networks and thermodynamics) of the proposed approach. \hyperref[sec:eTANN]{Section~\ref*{sec:eTANN}} presents the scheme of evolution Thermodynamics-based Artificial Neural Networks. In particular, in \hyperref[subsec:internalvariableknown]{Subsection~\ref*{subsec:internalvariableknown}}, we discuss the case of a priori known internal variables, while, in \hyperref[subsec:internalvariableunknown]{Subsection~\ref*{subsec:internalvariableunknown}}, the data-driven identification of the internal variables and their evolution equations is presented. \hyperref[subsec:inference]{Subsection~\ref*{subsec:inference}} presents the description of the material response in time, by means of the time integration of the evolution equations, performed outside of the network. In all these sections, particular attention is paid to juxtapose and show how the various general notions of solid mechanics are implemented in \textit{e}TANN. Finally, \hyperref[sec:applications]{Section~\ref*{sec:applications}} demonstrates the wide applicability of the proposed approach and its advantages with respect to previous models via several applications. First, considering the constitutive modeling at the material point level, we leverage behaviors displaying common features shared by the largest spectrum of materials: plasticity, damage, and viscosity. The approach is applied, then, to heterogeneous materials and, in particular, metamaterials displaying an elasto-viscoplastic behavior. At last, the proposed approach is used to perform multiscale simulations relying on the FEM$\times$TANN approach \citep[see][]{masistefanou2021}.\\

\section{Theoretical setting}
\label{sec:thermodynamics}
\noindent The following notation is used throughout the manuscript: $\bm{a}\cdot \bm{b} = a_i b_i$, $\bm{P}:\dot{\bm{F}}=P_{ij} \dot{F}_{ij}$, and $\text{div}\,\bm{y} = \frac{\partial y_i}{\partial x_j}$, with $\bm{x}$ the spatial coordinates, $i,j=1,2,3$. Einstein's summation is implied for repeated indices. 

\subsection{Neural networks in a nutshell}
\label{subsec:networks}
\noindent A feed-forward artificial neural network of depth $L$ has an input layer, $L-1$ hidden layers and an output layer ($L$-th layer). Data flows from layer $(k - 1)$ to layer $(k)$, with $k\in[1,L]$, according to
\begin{equation}
\bm{y}^{(k+1)} = \mathcal{A}^{(k)}\left(\bm{W}^{(k)} \cdot \bm{y}^{(k-1)} + \bm{b}^{(k)} \right),
\end{equation}
where $\bm{y}^{(k)}\in \mathbb{R}^{n}$ and $\bm{y}^{(k-1)}\in \mathbb{R}^{l}$; $\bm{W}^{(k)}\in \mathbb{R}^{n\times l}$ is the weights matrix between the $l$ nodes in layer $(k)$ and the $n$ nodes in layer $(k-1)$; $\bm{b}^{(k)}\in \mathbb{R}^{n}$ is the bias vector; and $\mathcal{A}^{(k)}$ is a nonlinear activation function (except for the case $k=L$, output layer, where the identity function is usually preferred).\\

During training, weights and biases are adjusted, in an iterative procedure \cite[see gradient descent algorithm and variations,][]{geron2019hands}, to minimize the error between the outputs (or predictions), $\bm{o}$, and benchmarks, $\bar{\bm{o}}$, measured by a loss function, $\mathcal{L}$, i.e.
\begin{equation}
\mathcal{L}(\bm{\bm{o}}) = \frac{1}{B} \sum_{i=1}^{B} \bm{\lambda} \cdot \rVert \bar{\bm{o}}^i -\bm{o}^i\lVert,
\label{eq:loss}
\end{equation}
where $B$ represents the size of the batch, $\bm{\lambda}$ are the hyperparameters determining the relative weighting of the terms of the loss function, and $\rVert \cdot \lVert$ denotes the following pseudo $l_1$-$l_2$ norm
\begin{equation}
 \rVert \bm{x}\lVert=\begin{cases}
|\bm{x}|-\frac{\bm{\delta}}{2} \quad \text{if } |\bm{x}|>\bm{\delta},\\
\frac{\bm{x}^2}{2\bm{\delta}} \quad \text{else.}
\end{cases}
\end{equation}
The training loss is backpropagated through each node to measure the error contribution from each connection and update the hyperparameters (\textit{reverse-mode differentiation}). This is done by computing the gradients of the loss with respect to the hyperparameters, by means of automatic differentiation. Once the training is completed, the neural network is used to make predictions from new inputs, at inference.
\subsection{Thermodynamic processes and constitutive restrictions}
\noindent We briefly recall the general continuum thermodynamic setting. We start by considering a body with reference configuration $\mathcal{B}$. The local form of the energy balance and dissipation rate inequality, under thermo-mechanic processes, in local form and referential description, reads
\begin{align}
&\dot{e} +\text{div}\bm{q} = \bm{P}:\dot{\bm{F}}+ \bm{r},\\
&\dbar = \dot{\eta} - \left[\theta^{-1}\bm{r}-\text{div} \left(\theta^{-1}\bm{q}\right)\right]\geq 0,
\end{align}
where $e$ and $\dot{e}$ are the volume density of the internal energy and its local time derivative; $\bm{P}$ is the first Piola-Kirchhoff stress tensor; $\bm{F}$ is the deformation gradient and $\dot{\bm{F}}$ its rate of change; $\bm{q}$ is the referential heat flux vector; $\bm{r}$ is the volume density of possible external source terms; $\dbar$ is the volume density of the total (mechanical plus thermal) dissipation rate; $\eta$ the volume density of entropy; and $\theta$ the absolute temperature. Combination of the first and second laws leads to the Clausius-Duhem inequality
\begin{equation}
\begin{split}
\dbar &= \bm{P}:\dot{\bm{F}} - \dot{e} +\theta\dot{\eta} - \theta^{-1} \bm{q}\cdot \bm{m}\\
&=\bm{P}:\dot{\bm{F}} - \left( \dot{\psi} +\dot{\theta}\eta \right) -\theta^{-1} \bm{q}\cdot \bm{m} \geq 0
\end{split}
\label{eq:thermo2}
\end{equation}
the latter involving the Helmholtz free energy density (per unit volume), $\psi=e-\theta s$, and denoting with $\bm{m}$ the temperature gradient.

\noindent By further introducing the internal dissipation rate $d$, whenever $\bm{m}=0$,  every admissible thermodynamic process in $\mathcal{B}$ must obey the Clausius-Duhem inequality
\begin{equation}
d = \bm{P}:\dot{\bm{F}} - \dot{\psi} -\dot{\theta}\eta \geq 0,
\label{eq:thermo3}
\end{equation}
at each time $t$ and for all material points $x\in\mathcal{B}$.

\noindent Constitutive restrictions on material processes can be derived from the Clausius-Duhem inequality (\ref{eq:thermo3}). Here, we follow the seminal work of \cite{coleman1967thermodynamics} and introduce internal state variables $z_j$, with $j=1,2,\ldots,n_{\bm{z}}$--where $n_{\bm{z}}$ is the (finite) number of internal state variables. Internal state variables or, internal variables for short, account for the influence of micromechanisms on the constitutive behavior of materials.\\
The free energy density can be thus regarded as function of the internal variables and the state variables, $\bm{F}$ and $\theta$--that is,
\begin{equation}
\psi = \hat{\psi}\left(\theta,\bm{F},\bm{z}\right),
\label{eq:energy}
\end{equation}
where the superposed caret in $\hat{\psi}$ serves to distinguish the free energy function from its values and $\bm{z}$ denotes the internal state vector $\bm{z}=\left(z_1,z_2, \ldots, z_{n_{\bm{z}}}\right)$. Alternatively, one may consider the elastic part of the deformation gradient as a state variable \citep[see e.g.][]{rubin2001physical,einav2012unification,dafalias2022split}. However, here we don't follow this second path as our target is to develop a data-driven constitutive modeling approach that is general enough to describe the largest range of materials, without requiring the decomposition of the deformation gradient, or its rate of change, into an elastic and a plastic part.\\
From Eq. (\ref{eq:energy}) follows that
\begin{equation}
\dot{\psi} = \partial_{\bm{F}}\hat{\psi} :\dot{\bm{F}}+ \partial_{\theta}\hat{\psi}\, \dot{\theta}+ \partial_{\bm{z}}\hat{\psi} \cdot\dot{\bm{z}},
\label{eq:energy_diff}
\end{equation}
where $\partial_{\bm{F}}\hat{\psi} = \sfrac{\partial \hat{\psi}}{\partial \bm{F}}$.
Substituting (\ref{eq:energy_diff}) into (\ref{eq:thermo3}) we obtain the following expression for the internal dissipation rate
\begin{equation}
d = (\bm{P}-\partial_{\bm{F}}\hat{\psi}):\dot{\bm{F}}-(\eta+\partial_{\theta}\hat{\psi})\,\dot{\theta}-\partial_{\bm{z}}\hat{\psi}\left(\theta,\bm{F},\bm{z}\right) \cdot\dot{\bm{z}}, \quad \forall \;\; \dot{\theta}, \dot{\bm{F}}, \dot{\bm{z}}
\end{equation}
with $-\partial_{\bm{z}}\hat{\psi}\left(\theta,\bm{F},\bm{z}\right)$ being the thermodynamic forces. It follows that every admissible thermodynamic process must satisfy the constitutive restrictions hereinafter, holding true at any time $t$,
\begin{align}
\label{eq:stress}
\bm{P} &= \partial_{\bm{F}}\hat{\psi}\left(\theta,\bm{F},\bm{z}\right),\\
\eta &= -\partial_{\theta}\hat{\psi}\left(\theta,\bm{F},\bm{z}\right),\\
d &= -\partial_{\bm{z}}\hat{\psi}\left(\theta,\bm{F},\bm{z}\right)\cdot\dot{\bm{z}}\geq 0.
\label{eq:dissipation}
\end{align}
A last ingredient allows the closure of the constitutive relationship: the evolution equation for the internal variables which has the general form
\begin{equation}
\dot{\bm{z}} = \bm{f}\left(\theta, \bm{F}, \bm{z}\right)
\label{eq:evolution}
\end{equation}
such that the dissipation inequality (\ref{eq:dissipation}) holds true. This form is justified by virtue of Truesdell's equipresence axiom \citep[see e.g.][]{truesdell1960modern}.\\
Note that, in some particular cases, e.g. when postulating the existence of a dissipation potential, analytical expressions for $\bm{f}$ satisfying the dissipation inequality by construction, can be retrieved \citep[see e.g.][]{maugin1994thermodynamicsA,einav2007coupled}. Similarly, evolution equations could be formulated as Onsagerian conductivity equations, i.e., $\dot{\bm{z}} = \bm{L} \cdot \partial_{\bm{z}}\hat{\psi}\left(\theta,\bm{F},\bm{z}\right)$ with $\bm{L}$ the so-called conductivity (positive definite) matrix, satisfying the dissipation inequality \citep[see][among others]{gurtin1996generalized,van2003weakly,einav2018hydrodynamic}. Yet, the above structure does not originate directly from thermodynamic considerations and may be not general enough, as it requires ad-hoc assumptions on the form of the internal variables rates and the thermodynamic forces. Thus, in the following, we opt for the more general expression given in Eq. (\ref{eq:evolution}). 

\section{Evolution TANN and the identification of evolution equations}
\label{sec:eTANN}
\noindent The thermodynamic framework presented above can be employed in the architecture of NN to learn physical and generalizable constitutive models from off-line and/or on-line streaming data, either derived from experiments or fine-scale simulations. Such neural networks, previously developed by the authors and co-workers \citep{masi2021thermodynamics,masistefanou2021}, are the so-called Thermodynamics-based Artificial Neural Networks (TANN). \\
To describe the time evolution of the constitutive response, TANN were originally developed relying on an incremental formulation, i.e. they were discrete-time. This allowed to identify internal variable increments corresponding to given values of the state variables and their increments. Yet, mixing time integration and constitutive modeling of materials diminishes the rigor and efficiency of the theoretical setting presented in \hyperref[sec:thermodynamics]{Section~\ref*{sec:thermodynamics}}. Motivated by the above necessity, we propose a new approach that we define as evolution TANN (\textit{e}TANN) allowing to describe material constitutive behaviors without the need of accounting for time and state variable increments. This is obtained by adopting a differential formulation, rather than an incremental one: all quantities of interest and their rates refer at time $t$. Although \textit{e}TANN share with the previous TANN some key features of their architecture, the former allow to learn the unknown evolution equations of the internal variables.

Depending on whether the latter are a priori known or not, there exist two possible configurations of \textit{e}TANN, as detailed below.
\subsection{Data-driven identification of evolution equations for a priori known internal variables}
\label{subsec:internalvariableknown}
\noindent We first consider the case where the internal variables of the material and their rates are known. Note that in absence of the latter, their finite difference approximation can be equivalently considered.

\begin{figure*}[ht]
	\centering
	\includegraphics[width=0.4\linewidth]{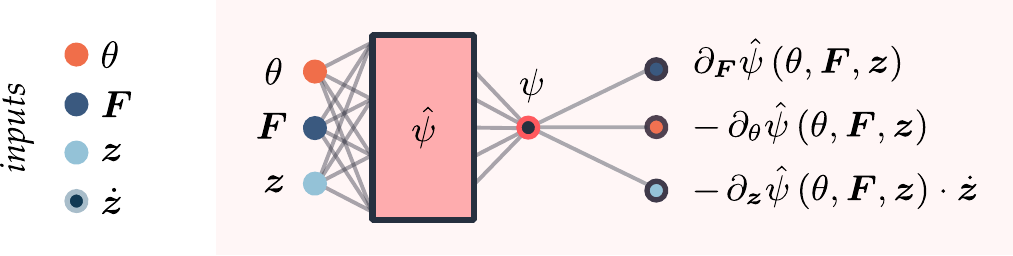}
\caption{Thermodynamics block. All quantities refer at time $t$.}
\label{fig:archtectureTANN}
\end{figure*}

\textit{e}TANN are composed of two building blocks: the free energy density network and the evolution equation network. The free energy network, shown in \hyperref[fig:archtectureTANN]{Figure~\ref*{fig:archtectureTANN}}, is responsible of the prediction of the material stress, from the fulfillment of the thermodynamics restrictions (\ref{eq:stress}-\ref{eq:dissipation}). It consists of one NN trained to predict the value of the free energy density--that is, $\psi=\hat{\psi}\left(\bm{F},\theta,\bm{z}\right)$, Eq. (\ref{eq:energy}). Note that we denote, with a slight abuse of notation, the neural network as the same function that is supposed to represent, after training. This formalism is kept within the entire manuscript.\\
Material stress, entropy, and internal dissipation rate are computed, using the constitutive restrictions (\ref{eq:stress}-\ref{eq:dissipation}), by relying on the automatic differentiation of the operator $\hat{\psi}$ with respect to its inputs.

The free energy network is trained by minimizing a loss composed of three contributions, and namely
\begin{equation}
\mathcal{L}_{\text{energy}} =\mathcal{L}_\psi +\mathcal{L}_{\text{grad} \psi}+\lambda_{\text{reg}}\mathcal{L}_\text{reg},
\end{equation}
where $\lambda_{\text{reg}}$ is a regularization weighting. The first term,
\begin{equation}
\mathcal{L}_\psi = \left\lVert \psi - \hat{\psi}\left(\theta,\bm{F},\bm{z}\right) \right\rVert,
\label{eq:loss_energy}
\end{equation}
ensures that the neural network can accurately predict the free energy density. The second term,
\begin{equation}
\mathcal{L}_{\text{grad} \psi} =\left\lVert \bm{P} - \partial_{\bm{F}} \hat{\psi}\left(\theta,\bm{F},\bm{z}\right)\right\rVert+
\left\lVert \eta + \partial_{\theta} \hat{\psi}\left(\theta,\bm{F},\bm{z}\right)\right\rVert+
\left \lVert d  + \partial_{\bm{z}} \hat{\psi}\left(\theta,\bm{F},\bm{z}\right)\cdot \dot{\bm{z}}\right \rVert,
\label{eq:loss_gradenergy}
\end{equation}
ensures that the gradients of the neural network $\hat{\psi}\left(\theta,\bm{F},\bm{z}\right)$ coincide with the material stresses, entropy density, and intrinsic dissipation rate. Finally, the last term,
\begin{equation}
\mathcal{L}_\text{reg} = \left \lVert \{ \partial_{\bm{z}} \hat{\psi} \left(\theta,\bm{F},\bm{z}\right)\cdot \dot{\bm{z}}\} \right \rVert,
\label{eq:loss_regularization}
\end{equation}
with $\{\cdot\}$ being the Macaulay brackets, acts as a regularization term penalizing the prediction of non positive-definite dissipation rates, thus enhancing the fulfillment of the dissipation inequality (second law).\\
Note that the loss associated with the free energy density, $\mathcal{L}_\psi$, is not a necessary condition for the fulfillment of the laws of thermodynamics and the accurate prediction of the material response. Indeed, it can be easily proved that, in absence of the aforementioned loss term, the free energy density network, trained only through its gradients, will still respect the restrictions (\ref{eq:stress}-\ref{eq:dissipation}).\\

\begin{figure}[h]
	\centering
	\includegraphics[width=0.25\textwidth]{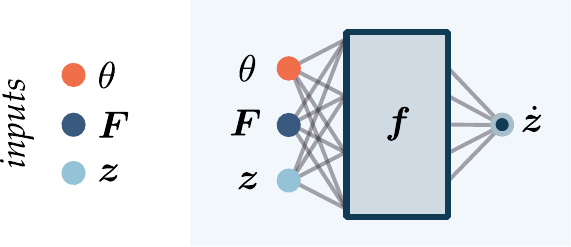}
	\caption{Evolution equation block for a priori determined internal variables. All quantities refer at time $t$.}
	\label{fig:archtectureTANN2}
\end{figure}

The evolution equation network, shown in \hyperref[fig:archtectureTANN2]{Figure~\ref*{fig:archtectureTANN2}}, is responsible for learning the evolution equations of the internal variables from the datasets collecting the the value of the state variables and their rates--that is, the function $\bm{f}\left(\theta, \bm{F}, \bm{z}\right)$. The training is performed by minimizing the error between the outputs and the rates of change of the internal variables,
\begin{equation}
\mathcal{L}_{\dot{\bm{z}}} = \left \lVert \dot{\bm{z}} -\bm{f} \left(\theta,\bm{F},\bm{z}\right) \right \rVert.
\end{equation}
Following the terminology in \citep{melchers2022comparison}, this is a derivative fitting approach. We opt for this approach as it results in a continuous-time form for the evolution equation and is computationally much cheaper compared to the trajectory fitting approach \citep[see][]{melchers2022comparison}.\\

Contrary to the classical TANN, the evolution NN learn the evolution equations, valid at any time $t$, rather than predicting the internal variable increments corresponding to a given time step and deformation increments. This is a significant difference with the previous formalism. 

The free energy and evolution equation networks can either be trained individually or together. In the latter case, the total loss function is given by the weighted sum of $\mathcal{L}_{\text{energy}}$ with $\mathcal{L}_{\dot{\bm{z}}}$. At inference, after training, the two networks are joined together to obtain a complete and thermodynamic consistent constitutive representation of the material state, at time $t$. This is achieved by means of the time integration of the evolution equations, as detailed in \hyperref[subsec:inference]{Subsection~\ref*{subsec:inference}}.

In the absence of rate dependency of the response, the proposed, new formulation gives accurate predictions that are independent of the increment size. Notice that the sensitivity of the response on the size of the increment is a major limitation of all existing approaches \citep[see][among others]{lefik2003artificial,lefik2009artificial,xu2020learning,vlassis2021sobolev,bonatti2021one,jones2022neural}. In parallel, the proposed approach is also able to accurately describe the behavior of complex materials with rate effects, thanks to the decoupling of the material behavior from the incrementation scheme.

\subsection{Data-driven identification of internal variables and evolution equations}
\label{subsec:internalvariableunknown}
\noindent The above scheme of \textit{e}TANN builds upon the knowledge of the internal variables. However, the crucial question is: \textit{how internal variables can be identified?}

Numerous approaches and applications demonstrated that the best way to determine the number and nature of internal variables is to discern the internal mechanisms and phenomena that are thought to influence the material behavior from those that are not (and thus pick out quantities that describe them). This is what Maugin defined as the \textit{Art of modelling} \citep{maugin2015saga}. Reformulating, the identification of internal variables is the responsibility of the modeler that needs to adapt, every time, to the specific application.

Whilst this point of view has been, and still is, widely employed, there is an emerging desire of conceiving a general framework able to uncover the internal variables for any material. Towards this direction, few works can be found. Without being exhaustive, we refer to \citet{grmela1997dynamics,van2008internal,berezovski2011thermoelasticity,hernandez2021deep,masistefanou2021}.\\

Here, we adopt the approach first developed in \cite{masistefanou2021}, where the identification of internal variables stems from the knowledge of the internal coordinates $\bm{\xi}$. These internal coordinates represent all those variables describing the material behavior at the microscopic scale. Thus, they gather microscopic representations of the degrees of freedom of the internal material structure, such as displacement, velocity, momentum fields and internal force networks. Internal variables can thus be identified by learning latent (low-rank) representations of the internal coordinates and enforcing the laws of thermodynamics. In other words, the approach postulates the existence of an operator $\bm{h}$,
\begin{equation}
\bm{z} = \bm{h} \left( \bm{\xi}\right) \qquad \text{such that}\quad  \psi = \hat{\psi}\left(\theta,\bm{F},\bm{z}\right).
\label{eq:operator1}
\end{equation} 
In parallel, without loss of generality, we postulate the existence of a second operator $\bm{g}$,
\begin{equation}
\tilde{\bm{\xi}} = \bm{g} \left(  \bm{h} \left( \bm{\xi}\right)\right) = \bm{g} \left( \bm{z}\right),
\label{eq:operator2}
\end{equation} 
where the superposed tilde serves to distinguish a low-rank approximation of the internal coordinates from the internal coordinates  themselves. The function $\bm{g}$ is said to be the pseudoinverse of $\bm{h}$.

In \citet{masistefanou2021}, the authors adopted the above formalism to determine internal variables from the microscopic fields of complex materials. However, the framework used was based on an incremental formulation \citep[discrete-time dynamics,][]{strogatz2018nonlinear}. Here we propose an extension, as part of the new framework of \textit{e}TANN, allowing for a suitable identification of the evolution equations of the internal variables in continuous-time, i.e., without relying on time discretization.\\
We start by noticing that the rate of change of internal variables--that is, the evolution equation--can be written as
\begin{equation}
\dot{\bm{z}} =  \dot{\bm{h}}(\bm{\xi}) = \partial_{\bm{\xi}} \bm{h}(\bm{\xi})\cdot \dot{\bm{\xi}},
\label{eq:evolution2}
\end{equation}
by differentiating in time Eq. (\ref{eq:operator1}) and considering $\bm{h} \left( \cdot\right)$ to be regular enough for the derivation to hold. Thus, the rate of change of the internal variables can be directly computed using Eq. (\ref{eq:evolution2}), rather than using an incremental formulation and without postulating a particular form of $\bm{f}$.\\
In a similar manner, the evolution of the internal coordinates can be expressed as
\begin{equation}
\dot{\tilde{\bm{\xi}}} =  \dot{\bm{g}}(\bm{z}) = \partial_{\bm{z}} \bm{g}(\bm{z})\cdot \dot{\bm{z}},
 \label{eq:evolution3}
\end{equation}
obtained by differentiating Eq. (\ref{eq:operator2}) and considering $\bm{g} \left( \cdot\right)$ to be regular enough for the derivation to hold. \\
It is clear how, once the operators $\bm{h}$ and $\bm{g}$ as well as the value of the internal coordinate rates are known, one can compute easily the rates of the internal variables using Eq. (\ref{eq:evolution2}). Once the rates are computed, we can use the evolution equation network to identify the operator $\bm{f}(\theta,\bm{F},\bm{z})$, thus determining the evolution equations of the internal variables.

At the heart of the identification of the internal variables and their evolution equations lies the search for an appropriate form of the functions $\bm{g}$ and $\bm{h}$. Here, we identify the latter as the neural networks of a standard autoencoder, being the operator $\bm{h}(\bm{\xi})$ the encoder and the operator $\bm{g}(\bm{z})$ the decoder. It can be proved that both operators are differentiable if the activation functions used are also differentiable, thus equations (\ref{eq:evolution2}) and (\ref{eq:evolution3}) hold true.\\
While autoencoders allow the reconstruction of the internal coordinates and their rates, a full reconstruction is not needed to characterize the material response. In addition, one may prefer other dimensionality reduction techniques, e.g. principal component analysis, among others \citep{geron2019hands,giovanniCFM}. The proposed approach is general and independent on the particular choice made to identify the latent representations of the internal coordinates.\\

\begin{figure*}[ht]
	\centering
	\includegraphics[width=0.60\textwidth]{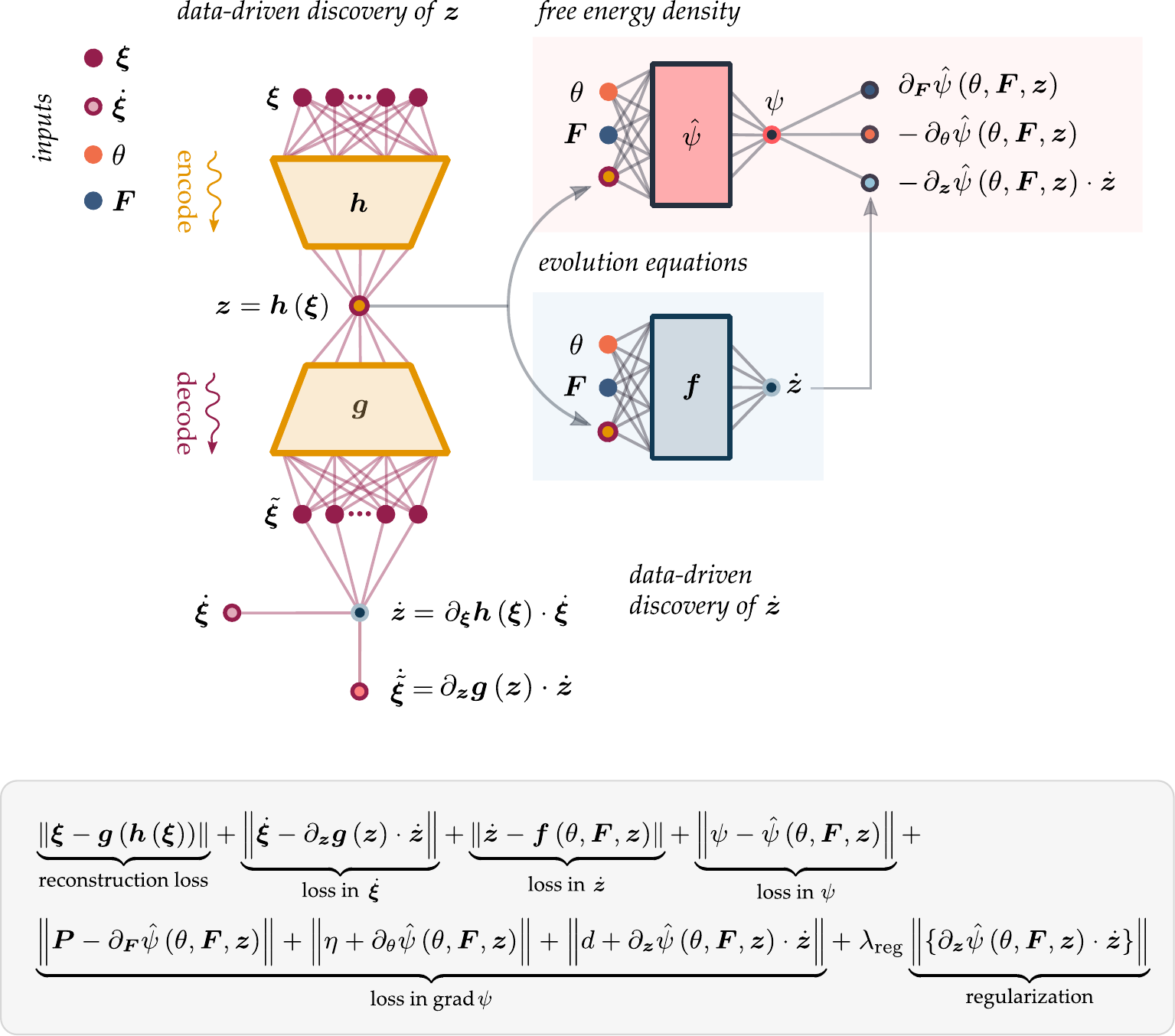}
	\caption{Evolution TANN for the data-driven identification of internal variables and governing equations: architecture for training (top) and minimized training loss (bottom). All quantities refer at time $t$. The internal variables are identified as the latent representations of the encoder, i.e., $\bm{z} = \bm{h} \left(\bm{\xi}\right)$. The internal variables rate is computed by means of the autodifferentiation of the encoder with respect to the internal coordinates, i.e., $\dot{\bm{z}} = \partial_{\bm{\xi}}\bm{h}(\bm{\xi})\cdot \dot{\bm{\xi}}$. }
	\label{fig:archtectureTANN_final}
\end{figure*}

\hyperref[fig:archtectureTANN_final]{Figure~\ref*{fig:archtectureTANN_final}} shows the architecture of \textit{e}TANN for the data-driven identification of the internal variables and evolution equations. The architecture is composed of one autoencoder, the free energy density network, and the evolution equation network. 
The identification of the internal variables is driven by the minimization of the reconstruction loss,
\begin{equation}
\mathcal{L}_\text{recon} = \left\lVert \bm{\xi} -\bm{g}\left( \bm{h} \left(\bm{\xi}\right)\right) \right\rVert,
\end{equation}
which ensures that the autoencoder can reconstruct the internal coordinates from the latent representations. Note that, at the end of the training, these latent representations, $\bm{h} \left(\bm{\xi}\right)$, will coincide with the internal variables, i.e., $\bm{z} \equiv \bm{h} \left(\bm{\xi}\right)$.\\
The identification of the evolution equations passes from the minimization of two functions. The first one,
\begin{equation}
\mathcal{L}_{\dot{\bm{\xi}}} = \left\lVert \dot{\bm{\xi}} - \partial_{\bm{z}}\bm{g}(\bm{z})\cdot \partial_{\bm{\xi}}\bm{h}(\bm{\xi})\cdot \dot{\bm{\xi}} \right\rVert,
\end{equation}
ensures the accurate computation of the rates of change of the internal coordinates using Eqs. (\ref{eq:evolution2},\ref{eq:evolution3}). In parallel, this allows to identify the internal variables rate simply as $\dot{\bm{z}}\equiv \partial_{\bm{\xi}}\bm{h}(\bm{\xi})\cdot \dot{\bm{\xi}}$. Once the value of $\dot{\bm{z}}$ is known, the second loss,
\begin{equation}
\mathcal{L}_{\dot{\bm{z}}} = \left\lVert \dot{\bm{z}}- \bm{f} \left(\theta,\bm{F},\bm{z}\right) \right\rVert,
\end{equation}
allows to learn the evolution equation--that is, the function $\bm{f} \left(\theta,\bm{F},\bm{z}\right)$. While standard autoencoders can be trained in isolation to discover dimensionality reductions, there is no guarantee that the latent representations will coincide with a thermodynamic consistent set of internal variables. Thus, in addition to the above losses, the loss in the free energy density $\mathcal{L}_\psi$ and its gradients $\mathcal{L}_{\text{grad}\psi}$--Eqs. (\ref{eq:loss_energy},\ref{eq:loss_gradenergy})--are needed. A regularization loss for the fulfillment of the dissipation inequality, $\mathcal{L}_\text{reg}$, is also used. The combination of the above terms gives the overall loss function
\begin{equation}
\mathcal{L} = \mathcal{L}_\text{recon}+\mathcal{L}_{\dot{\bm{\xi}}}+\mathcal{L}_{\dot{\bm{z}}}+\mathcal{L}_\psi+\mathcal{L}_{\text{grad}\psi}+\lambda_\text{reg}\mathcal{L}_\text{reg},
\label{eq:training_loss}
\end{equation}
where weighting hyperparameters, as defined in Eq. \ref{eq:loss}, are considered to be unitary, as all quantities of interest are normalized (see also \hyperref[sec:applications]{Section~\ref*{sec:applications}}).
It is worth noticing that, as for the case where the internal variables are a priori known, the training of the NN composing the \textit{e}TANN can also be performed individually. In this case, the autoencoder and the free energy density network can be first trained together, by minimization of the weighted sum of the losses $\mathcal{L}_\text{recon},\mathcal{L}_{\dot{\bm{\xi}}},\mathcal{L}_\psi,\mathcal{L}_{\text{grad}\psi}$, and $\mathcal{L}_\text{reg}$, then, the evolution equation network is trained by minimization of the loss $\mathcal{L}_{\dot{\bm{z}}}$. Both training procedures have been tested and deliver the same results.\\

It is worth noticing that expression (\ref{eq:evolution2}) only serves to identify the form of the evolution equations, i.e., $\bm{f}$, and it is not used herein as a substitute for it. Indeed, if we express the rate of change of the internal variables as function of the internal coordinates, then we will be obliged to carry the values of the coordinates, at any time, at inference. This would result in a cumbersome and time-consuming approach, rapidly made prohibitive by the available memory for the calculations. Instead, if Eq. (\ref{eq:evolution2}) is used to retrieve the appropriate representation of $\bm{f}$, during training, then only the evolution equation network and the state variables are required to characterize the material behavior.

In essence, the proposed framework of \textit{e}TANN allows not only to identify the evolution equations of the internal variables, but also to track the dynamics of the internal coordinates--that is, the microscopic material mechanisms and phenomena--at reduced computational cost.

\subsection{Inference mode and time integration}
\label{subsec:inference}
\begin{figure*}[hbt]
	\centering
	\includegraphics[width=0.725\textwidth]{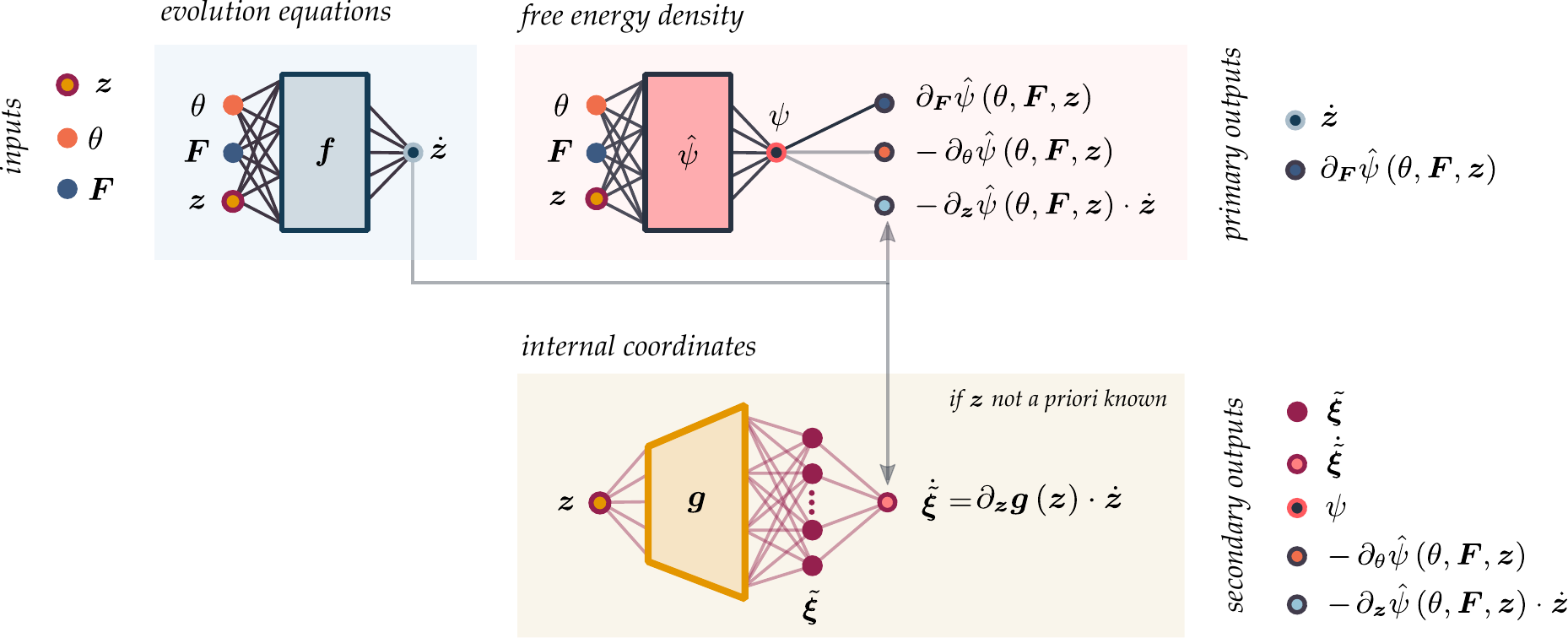}
	\caption{Structure of the evolution TANN at inference for both a priori known internal variables and identified internal variables. The network predicts the material stress and the internal variable rates to deliver a concise description of the material response (graph in bold). Additionally, the network can output secondary quantities (shaded graph), such as the free energy density and the dissipation rate, if needed. In the case where the internal variables are identified from the internal coordinates, the decoder can be kept to reconstruct the material microstructure and its evolution. }
	\label{fig:TANN_inference}
\end{figure*}

\noindent Once \textit{e}TANN are trained, they can be used in inference mode as classical constitutive models, i.e.
\begin{equation}
\dot{\bm{z}}(t),\bm{P}(t) = \bm{e}\textbf{TANN}\big(\theta(t), \bm{F}(t), \bm{z}(t)\big) \quad \forall\, t.
\label{eq:TANN}
\end{equation}
\hyperref[fig:TANN_inference]{Figure~\ref*{fig:TANN_inference}} shows the structure of \textit{e}TANN at inference, for both the cases where the internal variables are priori known and where they are not. The graph highlighted in bold is the minimal structure needed to predict the material response at time $t$. For the case of a priori known internal variables, only the evolution equation and free energy networks are representative. Instead, in the case where the internal variables are identified, these are conclusively identified as the latent representations of the encoder, which is removed completely from the architecture of the \textit{e}TANN. However, the decoder may be kept to map $\bm{z}$ back to the internal coordinates, i.e. $\tilde{\bm{\xi}}(t)=\bm{g}\left(\bm{z}(t)\right)$, and compute their rates, $\dot{\tilde{\bm{\xi}}}(t)=\partial_{\bm{z}}\bm{g}(\bm{z}(t))\cdot \bm{f}(\theta(t), \bm{F}(t), \bm{z}(t))$, if the particular application at hands requires it. In addition, the free energy, the entropy, and the dissipation rate densities can be computed.\\ 

The aim of obtaining constitutive models is to used the latter for calculating the material response under general loading paths. In displacement-based formulations, such as the Finite Element Method (FEM), loading paths are expressed in terms of strain and temperature loads. However, if one opts for a stress-based formulation, this can be obtained with the proposed approach by simply selecting the Gibbs free energy density instead of Helmholtz free energy density (\ref{eq:energy}).\\
Independently of the formulation, an incremental form of the constitutive relationship is required. This turns to be straightforward with the new formalism of \textit{e}TANN. Once the material state has been calculated at time $t$, using (\ref{eq:TANN}), we simply need to integrate the evolution equations to obtain the updated value of the internal variables at the next increment. Thus, the problem reduces to the solution of the following initial value problem
\begin{equation}
\dot{\bm{z}}(t) = \bm{f}\big(\theta(t), \bm{F}(t), \bm{z}(t)\big), \quad \bm{z}(t_0)=\bm{z}_0.
\label{eq:ode}
\end{equation}
The above equation can be solved, for an unknown evolution of $\theta$ and $\bm{F}$, using the artillery of well-known and studied numerical integration methods. Alternatively, one may prefer using neural networks as time integrators \citep[see][among others]{lu2019deeponet,kutz2022parsimony}.\\
Herein, we will use the established implicit multi-step method based on a backward differentiation formula \citep{2020SciPy-NMeth}, which does not require the training of additional NN.

\section{Applications}
\label{sec:applications}
\noindent This Section provides several examples and applications of the proposed approach to a variety of homogeneous and heterogeneous materials, modeled via fine-scale simulations. First, we investigate the modeling at the material point level. In particular, we consider plasticity, damage, rate-dependency and combinations of them, being the main ingredients characterizing the behavior of a broad range of materials. Second, multiscale simulations, relying on the approach developed in \cite{masistefanou2021}, are presented to investigate the performances of \textit{e}TANN in accelerating state-of-the-art FE\textsuperscript{2} simulations.\\

\noindent In all applications, materials are assumed in isothermal conditions, i.e., $\theta(t)=\theta_0$ for all $t$, and small-strain regimes. Accordingly, we have
$\bm{\sigma}\approx \bm{P}$, $\bm{\varepsilon}\approx \bm{F}^{\text{Sym}} +\bm{I}$, with $\bm{\sigma}$ being the Cauchy stress tensor, $\bm{\varepsilon}$ the infinitesimal strain tensor, and $\bm{F}^{\text{Sym}}$ denoting the symmetric part of the deformation gradient. \\
Material datasets for the training of the networks are generated according to the procedure detailed in Appendix. In preprocessing, all quantities of the generated datasets are split into training (60\%), validation (20\%), and test (20\%) sets and normalized between -1 and 1. In particular, given a quantity $\bm{x}$, its normalized value ${\bm{x}}^{\sharp}$ is computed as 
\begin{equation}
{\bm{x}}^{\sharp} = \frac{1}{\bm{\alpha}_{\bm{x}}}\left(\bm{x}-\bm{\beta}_{\bm{x}}\right),  \quad \alpha_{x_i} = \frac{1}{2}\left( \max(x_i) - \min(x_i)\right),\quad \beta_{x_i} = \frac{1}{2}\left( \max(x_i)+ \min(x_i)\right)
\end{equation}
where $\bm{\alpha}_{\bm{x}}$ and $\bm{\beta}_{\bm{x}}$ are computed from the statistics of the training and validation sets.\\

\noindent From a physical perspective, in the applications herein considered, the free energy density and the rates of the internal variables are expected to vanish when the state variables are zero. This condition is strongly imposed by redefining the outputs of the free energy density and the evolution equation networks respectively as
\begin{equation*}
\psi = \alpha_{\psi}\left(\hat{\psi}^{\sharp}({\bm{\varepsilon}},{\bm{z}})- \hat{\psi}^{\sharp}({\bm{0}},{\bm{0}})\right),\quad
\dot{\bm{z}} = \bm{\alpha}_{\dot{\bm{z}}}\left( \bm{f}^{\sharp}({\bm{\varepsilon}},{\bm{z}})- \bm{f}^{\sharp}({\bm{0}},{\bm{0}})\right),
\end{equation*}
where $\hat{\psi}^{\sharp}({\bm{\varepsilon}},{\bm{z}})=\hat{\psi}({\bm{\varepsilon}}^{\sharp},{\bm{z}}^{\sharp})$ and $\bm{f}^{\sharp}({\bm{\varepsilon}},{\bm{z}})=\bm{f}({\bm{\varepsilon}}^{\sharp},{\bm{z}}^{\sharp})$. In the data-driven identification of internal variables, we shall assume, without loss of generality, that the internal variables are zero when $\bm{\xi}=0$, and viceversa. This is achieved by construction considering
\begin{equation*}
\bm{z} = \bm{\alpha}_{\bm{z}} \left( \bm{h}^{\sharp}({\bm{\xi}})-\bm{h}^{\sharp}({\bm{0}})\right),\quad
\tilde{\bm{\xi}} = \bm{\alpha}_{\bm{\xi}} \left( \bm{g}^{\sharp}({\bm{z}})-\bm{g}^{\sharp}({\bm{0}})\right),
\end{equation*}
where the scaling constant $\bm{\alpha}_{\bm{z}}$ can be either considered as a trainable hyperparameter in the gradient descend algorithm or computed, after training, on the statistics of the internal variables. The above allows to enforce $\bm{z}(t=0)=\bm{0}$ as initial condition in the solution of the initial value problem (\ref{eq:ode}).\\

\noindent For the architecture and training of \textit{e}TANN we leverage TensorFlow \citep{abadi2016tensorflow}. In all applications, the hyperparameters of the networks composing \textit{e}TANN (i.e., number of hidden layers, neurons, activation functions, etc.) are selected to obtain the smallest error on the test datasets. The learning rate is set to $10^{-3}$ and early-stopping \citep{geron2019hands} avoid eventual overfitting on the training datasets. We use Adam optimizer with Nesterov’s acceleration gradient, with decay parameter equal to $0.9999$, and a mini batch size of 1000 samples is used.

As far it concerns the training procedure, first we train the free energy density (and the autoencoder) network, then we identify the form of the evolution equations by training the evolution equation network individually.

\subsection{Applications at the material point level}
\subsubsection{Elasto-plastic materials}
\label{par:rateindependent}
\noindent The first example concerns rate-independent plasticity. Especially, we consider a three-dimensional homogeneous material and use the von Mises yield criterion with isotropic hardening.\\
The material does not possess any microstructure. The internal variables are hence known and chosen to be the inelastic deformations, $\bm{z}^{pl}$, and the deviatoric invariant of the accumulated inelastic deformations, $z^{q}$. Accordingly, \textit{e}TANN are composed of the free energy density network and the evolution equation network, see \hyperref[fig:archtectureTANN]{Figure~\ref*{fig:archtectureTANN}} and \hyperref[fig:archtectureTANN2]{Figure~\ref*{fig:archtectureTANN2}}.

\begin{figure*}[ht]
     \centering
     \includegraphics[width=0.8\textwidth]{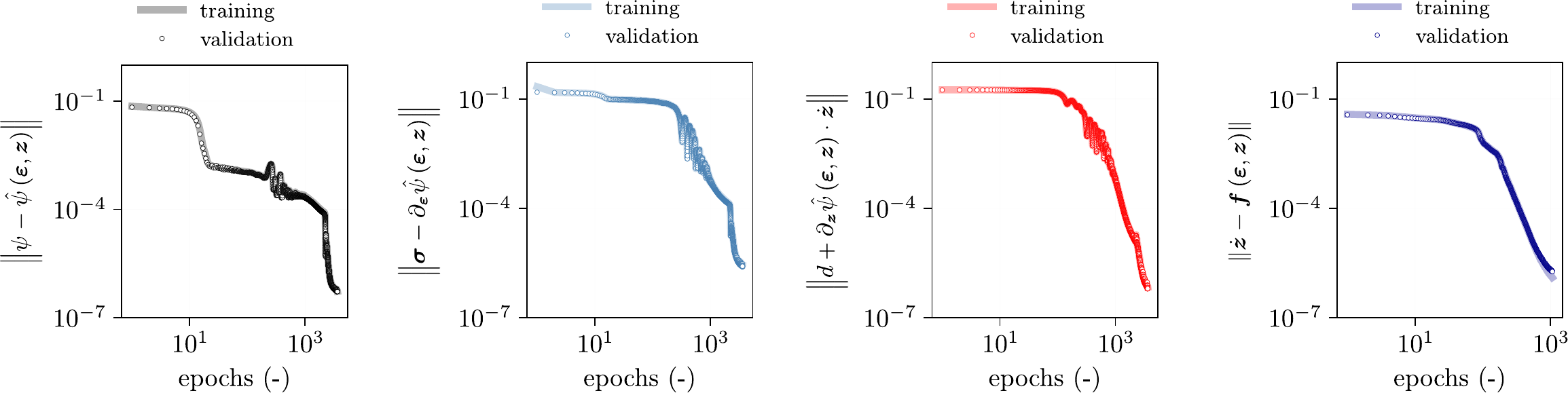}
        \caption{Training and validation losses of the free energy network (\textit{left}) and the evolution equation network (\textit{right}), during training.}
        \label{fig:training}
\end{figure*}

\subparagraph{Datasets, hyperparameters, and training.}
The material has the following parameters $K=167$ GPa, $G=77$ GPa, $c=100$ MPa, $H=10$ MPa, where $K$ and $G$ are the bulk and the shear modulus, $c$ is the material strength in simple shear, and $H$ the hardening modulus. 40'000 datasets are generated and used to train the network. The 80\textsuperscript{th} and 100\textsuperscript{th} percentiles of the imposed strains and strain rates are, respectively, $8\times 10^{-3}$ and $2\times 10^{-2}$, and, $3.4\times 10^{-4}$ and $2.4\times 10^{-3}$ s\textsuperscript{-1}. The free energy network consists of one hidden layer, with 48 nodes, and Gaussian error linear unit \citep[GELU, ][]{hendrycks2016gaussian} activation. The evolution equation network has five hidden layers, with 192 nodes each, and GELU activations. The total number of hyperparameters is 157'736. Convergence of the training loss is reached after approximately 3000 epochs, see \hyperref[fig:training]{Figure~\ref*{fig:training}}.

\subparagraph{Results.}
The predictions of \textit{e}TANN, at inference, are shown in \hyperref[fig:biaxial]{Figure~\ref*{fig:biaxial}} for a (unseen) cyclic triaxial loading path, with $\varepsilon_{22}=\varepsilon_{33}=-0.5\varepsilon_{11}$. Comparison with the reference model shows the high performance of the proposed approach. Note the ability of the network to identify the evolution of the yield strength as function of the internal variables. In parallel, the underlying thermodynamics is respected and the network successfully identifies the evolution equation of the internal variables.\\
\hyperref[fig:triaxial_hardening_error]{Figure~\ref*{fig:triaxial_hardening_error}} shows the relative absolute error in the stress predictions as function of the loading steps, whose mean value is found to be approximately $0.2\%$. This is a direct consequence of the generalization capabilities within the training set (interpolation) of \textit{e}TANN.

\begin{figure*}[ht]
	\centering
	\includegraphics[width=0.8\textwidth]{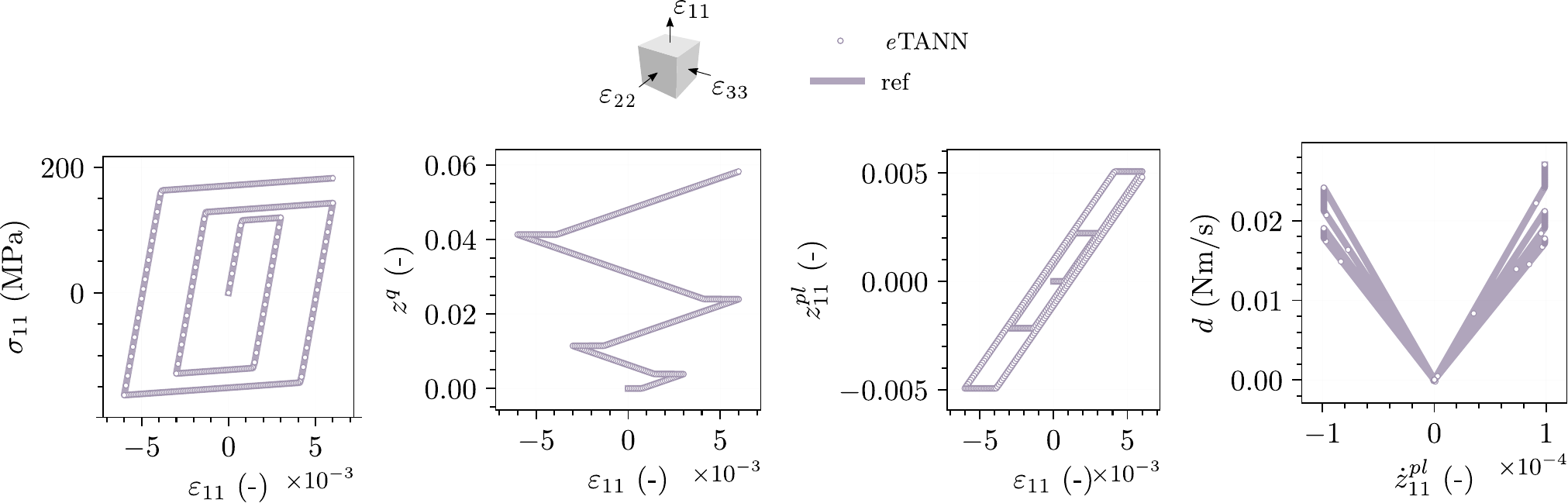}
	\caption{Triaxial loading path ($\varepsilon_{22}=\varepsilon_{33}=-0.5\varepsilon_{11}$) of a rate-independent elasto-plastic material with isotropic hardening, $\dot{\varepsilon}=10^{-4}$ s\textsuperscript{-1}.}
	\label{fig:triaxial_hardening}
\end{figure*}

\begin{figure*}[ht]
	\centering
	\includegraphics[width=0.6\textwidth]{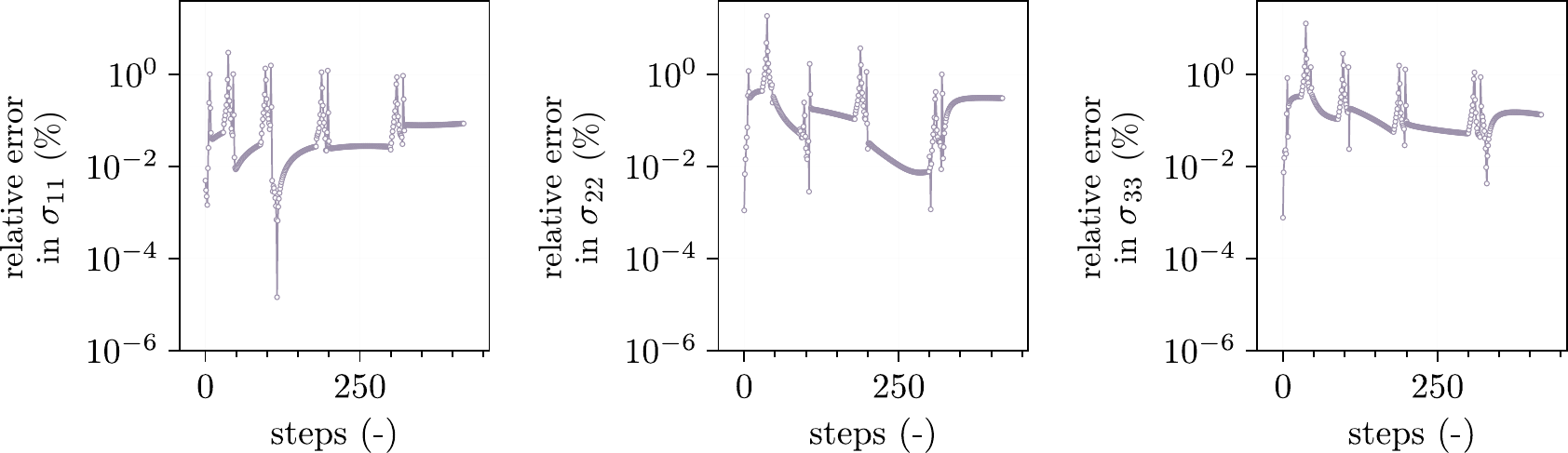}
	\caption{Relative absolute error in the stress prediction for the triaxial loading path in \hyperref[fig:triaxial_hardening]{Figure~\ref*{fig:triaxial_hardening}}.}
	\label{fig:triaxial_hardening_error}
\end{figure*}

\begin{figure*}[ht]
	\centering
	\includegraphics[width=0.8\textwidth]{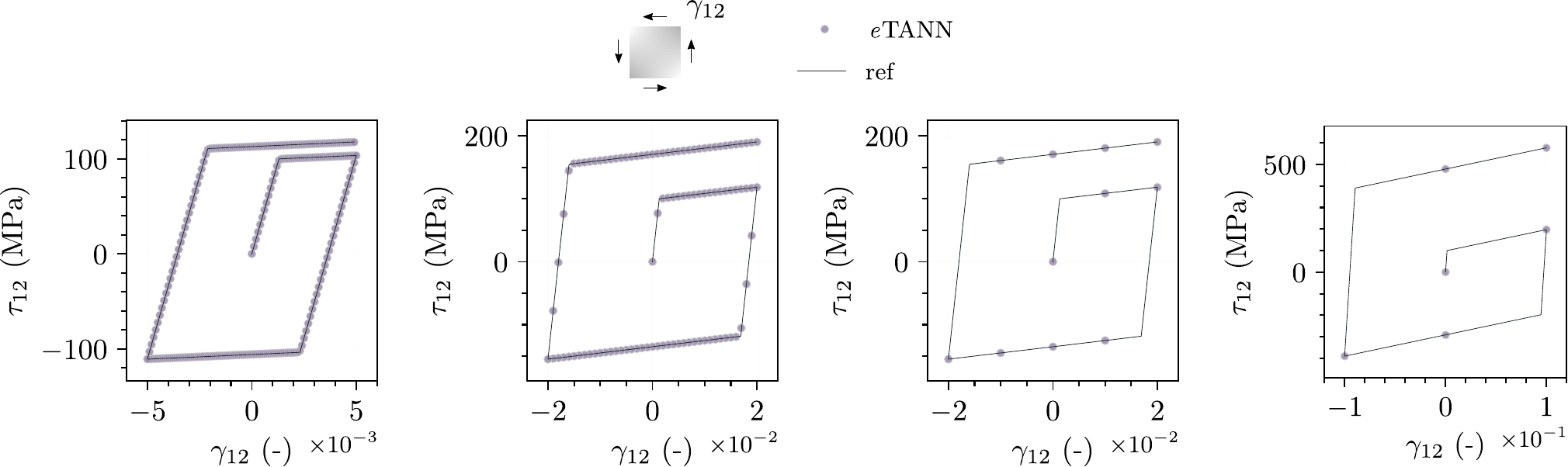}
	\caption{Simple shear loading path at varying of the strain increment ($\Delta \gamma$). From left to right: $\Delta \gamma=10^{-4}$, $10^{-3}$, $10^{-2}$, $10^{-1}$.}
	\label{fig:shear_hardening}
\end{figure*}

\noindent In contrast with the classical TANN, \textit{e}TANN deliver a differential representation of the evolution equations. The time integration of the the latter is carried outside of the network. Thus, the predictions are always independent of the values of strain increment and time step, in contrast with all existing approaches. This is clearly shown in \hyperref[fig:shear_hardening]{Figure~\ref*{fig:shear_hardening}}, where the predictions of the network are compared with the reference model for a simple shear path, applied through strain increments of different amplitude. For each of the loading paths, \textit{e}TANN predict with good accuracy the material response.\\
Another interesting result is the generalization of the network (and the constitutive model) outside of the training set. This is presented in \hyperref[fig:shear_hardening]{Figure~\ref*{fig:shear_hardening}}, for $\Delta \gamma=10^{-1}$, where the constitutive response is well predicted even for strain values that are far outside of the training set (approximately one order of magnitude).

\subsubsection{Elasto-plastic materials with damage}
\label{par:damage}

\noindent The second example consists of an elasto-plastic homogeneous material, with von Mises yield criterion and isotropic damage. Damage is modeled relying on the shear fracture model, a phenomenological model predicting the onset of damage due to shear band localization, as implemented in \citep{abaqus}. The damage initialization criterion depends on the value of the equivalent plastic strain, the shear stress ratio $\tau_s$, and a material parameter, $k_s$. At the onset of damage, softening of the yield strength and degradation of the stiffness take place following the evolution of a scalar damage variable, assumed exponential with respect to the plastic displacement up to a maximum value $z^d_f$.\\
The internal variables are selected to be the inelastic deformations, $\bm{z}^{pl}$, and the damage variable, $z^d$.

\subparagraph{Datasets, hyperparameters, and training}
The material has the following parameters $K=220$  GPa, $G=57$ GPa, $c=58$ MPa, $k_s=0.12$, $\tau_s = 0.94$,  $z^d_f=0.99$. The evolution of the damage variable is given by $z^d = [1-\exp(-a z_{eq}^{pl}/z_{eq,f}^{pl})]\left(1-\exp(-a)\right)$, where $z_{eq}^{pl}$ is the equivalent plastic strain, $a=3$ the decay parameter, and the ultimate equivalent plastic strain $z_{eq,f}^{pl}=0.004$.\\
40'000 datasets are generated and used to train the network. The 80\textsuperscript{th} and 100\textsuperscript{th} percentiles of the imposed strains and strain rates are, respectively, $1.2\times 10^{-3}$ and $6\times 10^{-3}$, and, $6.7\times 10^{-5}$ and $4.8\times 10^{-4}$ s\textsuperscript{-1}. The free energy network consists of three hidden layers, with 312 nodes each and GELUs activations. The evolution equation network has six hidden layers, with 192 nodes each, and GELU activations. The total number of hyperparameters is 390'404. Convergence of the training loss is reached after approximately 3000 epochs.

\begin{figure}[h]
  \centering
  \includegraphics[width=0.8\linewidth]{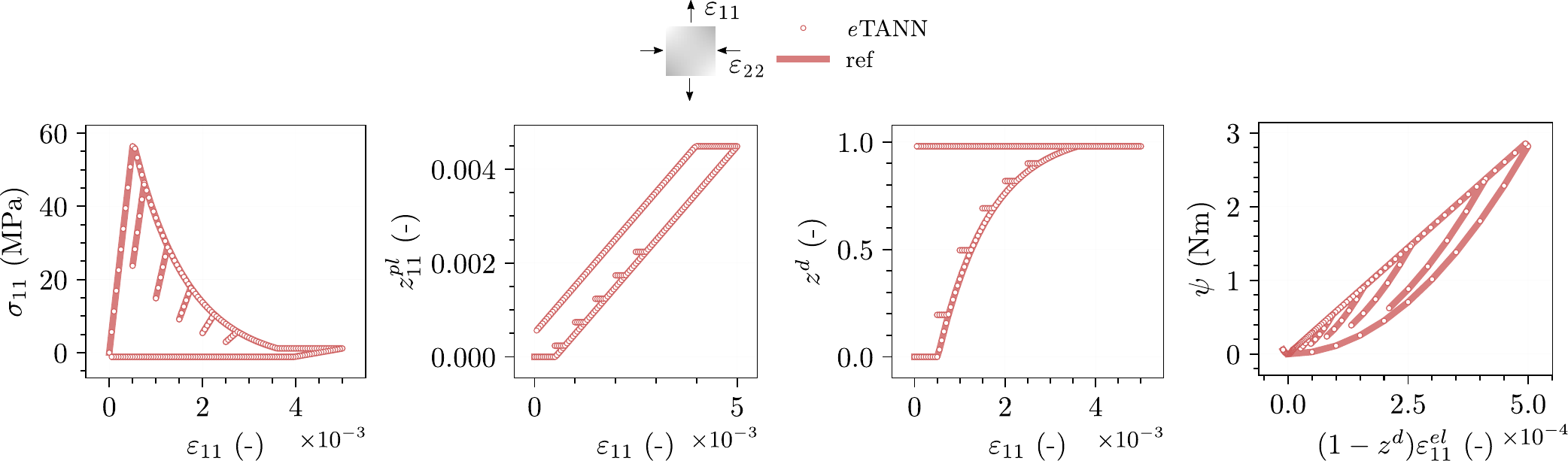}
	\caption{Biaxial loading path ($\varepsilon_{22}=-\varepsilon_{11}$).}
	\label{fig:damage_biaxial}
\end{figure}

\subparagraph{Results}
We show the predictions for a biaxial path, with multiple unloading and reloading, in \hyperref[fig:damage_biaxial]{Figure~\ref*{fig:damage_biaxial}}. \textit{e}TANN perfectly predict the material behavior and the evolution of the internal variables independently of the particular challenging loading path. Interesting is to observe the predictions of the damage scalar variable, $z_d$. Once full degradation of the material stiffness is reached, the model consistently predicts a constant value of the damage variable, which cannot exceed the maximum value $z^d_f = 0.99$.

\begin{figure}[h]
  \centering
  \includegraphics[width=0.6\linewidth]{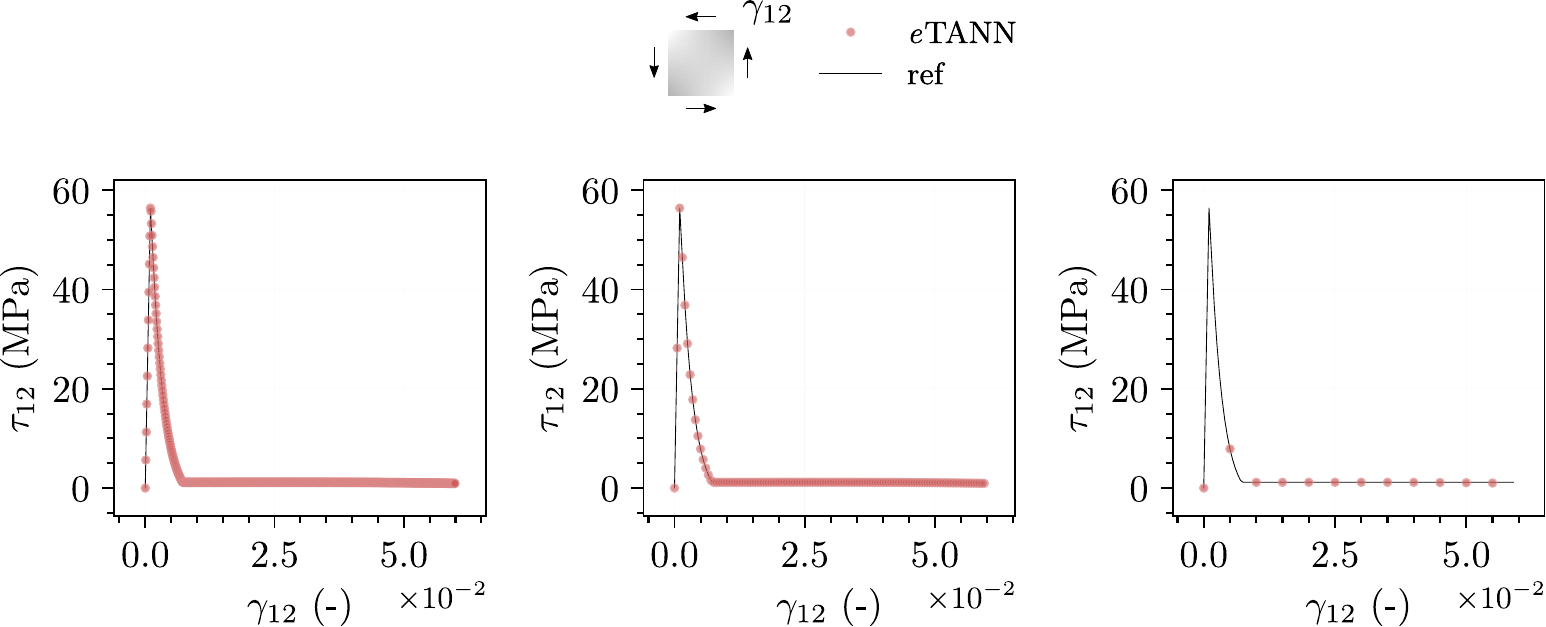}
	\caption{Simple shear loading path at varying of the strain increment ($\Delta \gamma_{12}$). From left to right: $\Delta \gamma_{12}= 10^{-4}$, $\Delta \gamma_{12}=5\times 10^{-4}$, $\Delta \gamma_{12}=5\times 10^{-3}$. }
	\label{fig:damage_shear}
\end{figure}

\noindent To illustrate the generalization of the \textit{e}TANN constitutive model, we consider a simple shear monotonous loading path, as shown in \hyperref[fig:damage_shear]{Figure~\ref*{fig:damage_shear}}. The predictions of the network and the integration of the evolution equations are independent of the particular strain increment (and time step). In addition, the network perfectly generalizes at strain values which are more than one order of magnitude larger than those in the training dataset (i.e., $6\times 10^{-3}$).

\subsubsection{Rate-dependent materials}
\label{par:viscosity}
\noindent In contrast with the previous applications, here we consider materials displaying rate effects. Especially, we consider a three-dimensional homogeneous material and use the von Mises yield criterion (perfect plasticity) and the Perzyna viscoplasticity model, see \cite{stathas2022role}.

\subparagraph{Datasets, hyperparameters, and training.}
The material has the following parameters $K=67$ GPa, $G=100$ GPa, $c=100$ MPa, $\mu=1$ s, where $\mu$ is the viscosity parameter. 60'000 datasets are generated and used to train the network. The 80\textsuperscript{th} and 100\textsuperscript{th} percentiles of the imposed strains and strain rates are $4.6\times 10^{-3}$ and $2\times 10^{-2}$, and, $33$ and $150$ s\textsuperscript{-1}, respectively. The hyperparameters of \textit{e}TANN are kept the same as for the case of the rate-independent case, \hyperref[par:rateindependent]{Paragraph~\ref*{par:rateindependent}}.

\begin{figure}[h]
	\centering
  \includegraphics[width=0.8\linewidth]{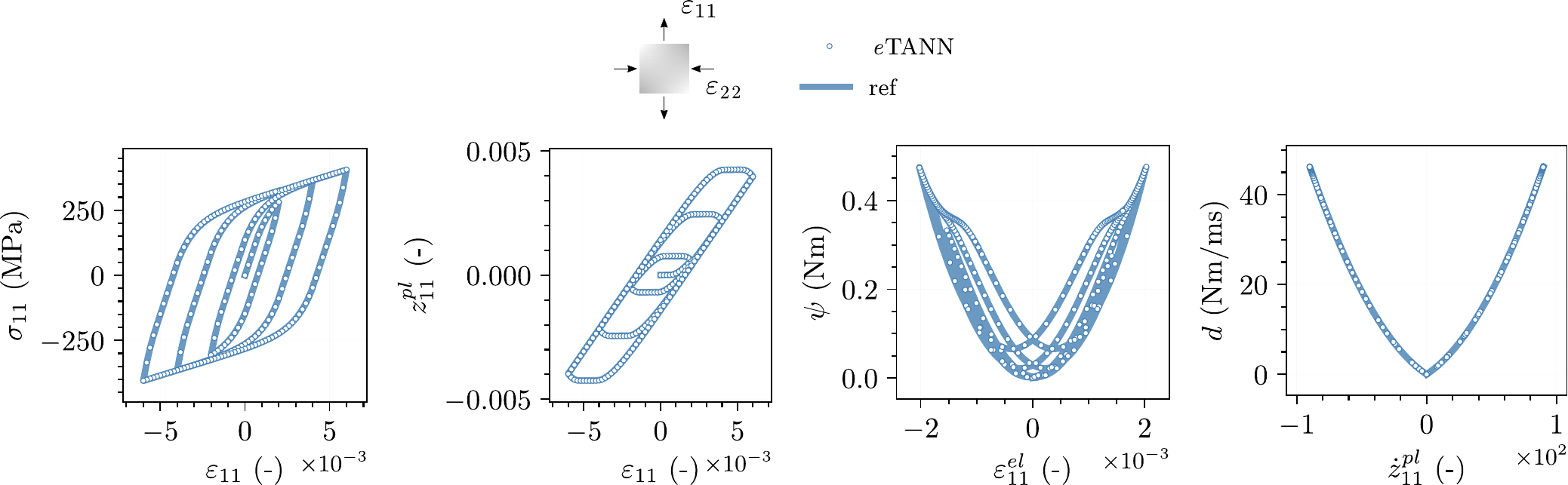}
	\caption{\small Predictions of the network for a biaxial loading path ($\varepsilon_{22}=-0.7\varepsilon_{11}$, $\dot{\varepsilon}=100$ s\textsuperscript{-1}) compared to the reference model. From left to right: stress versus total deformation, plastic ($\varepsilon^{pl}$) versus elastic ($\varepsilon^e$) deformations, free energy versus elastic deformation, and dissipation rate versus plastic strain rate ($\dot{\varepsilon}^{pl}$). }
	\label{fig:biaxial}
\end{figure}

\begin{figure*}[ht]
	\centering
	\includegraphics[width=0.6\textwidth]{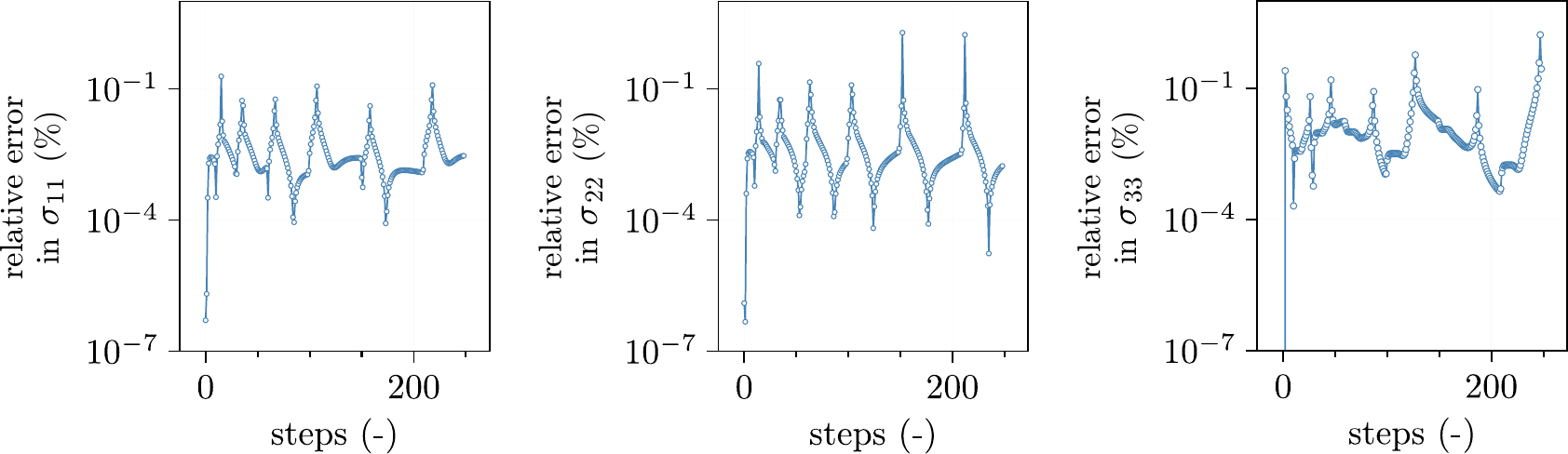}
	\caption{Relative error in the stress prediction versus the step increments, for the biaxial loading path in \hyperref[fig:biaxial]{Figure~\ref*{fig:biaxial}}.}
	\label{fig:biaxial_error}
\end{figure*}

\subparagraph{Results.}
The results of the network are shown in \hyperref[fig:biaxial]{Figure~\ref*{fig:biaxial}} for a biaxial loading path, with $\varepsilon_{22}=-0.7\varepsilon_{11}$ and $\dot{\varepsilon}=100$ s\textsuperscript{-1}. As for the previous applications, \textit{e}TANN is found to perfectly describe the material response, while satisfying the dissipation inequality, as well. The relative absolute errors in the stress prediction are shown in \hyperref[fig:biaxial_error]{Figure~\ref*{fig:biaxial_error}}. 

\begin{figure}[h]
	\centering
  \includegraphics[width=0.8\linewidth]{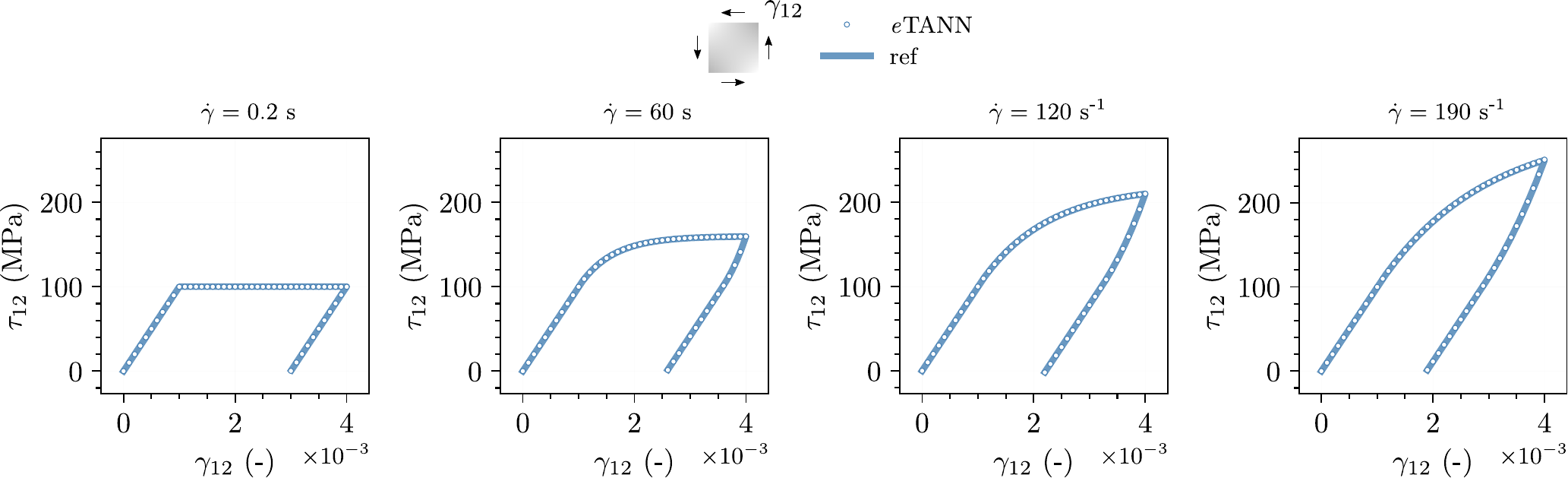}
	\caption{\small Simple shearing loading-unloading behavior at different strain rates of a viscoplastic material. From left to right: 0.2, 60, and 120, and 190 s\textsuperscript{-1}.}
	\label{fig:rates}
\end{figure}

\noindent Differently from the previous applications, rate dependency results in the dependence of the material stress with respect to the strain rate. This is accurately reproduced by the proposed approach as shown in \hyperref[fig:biaxial_error]{Figure~\ref*{fig:biaxial_error}} for a simple shear loading path at various strain rates. The model not only captures the increase of the material stress during loading, but also it accurately describes the material behavior at unloading. This shows how the proposed approach is scalable and general enough to predict a large variety of material behaviors. \\
In addition, \textit{e}TANN generalize well also outside of the training set. The last loading path in \hyperref[fig:rates]{Figure~\ref*{fig:rates}} refers to strain rates as high as 190 s\textsuperscript{-1}, which is far greater than maximum value in the training dataset (i.e., $150$ s\textsuperscript{-1}).

\begin{figure}[h]
  \centering
  \includegraphics[width=0.6\linewidth]{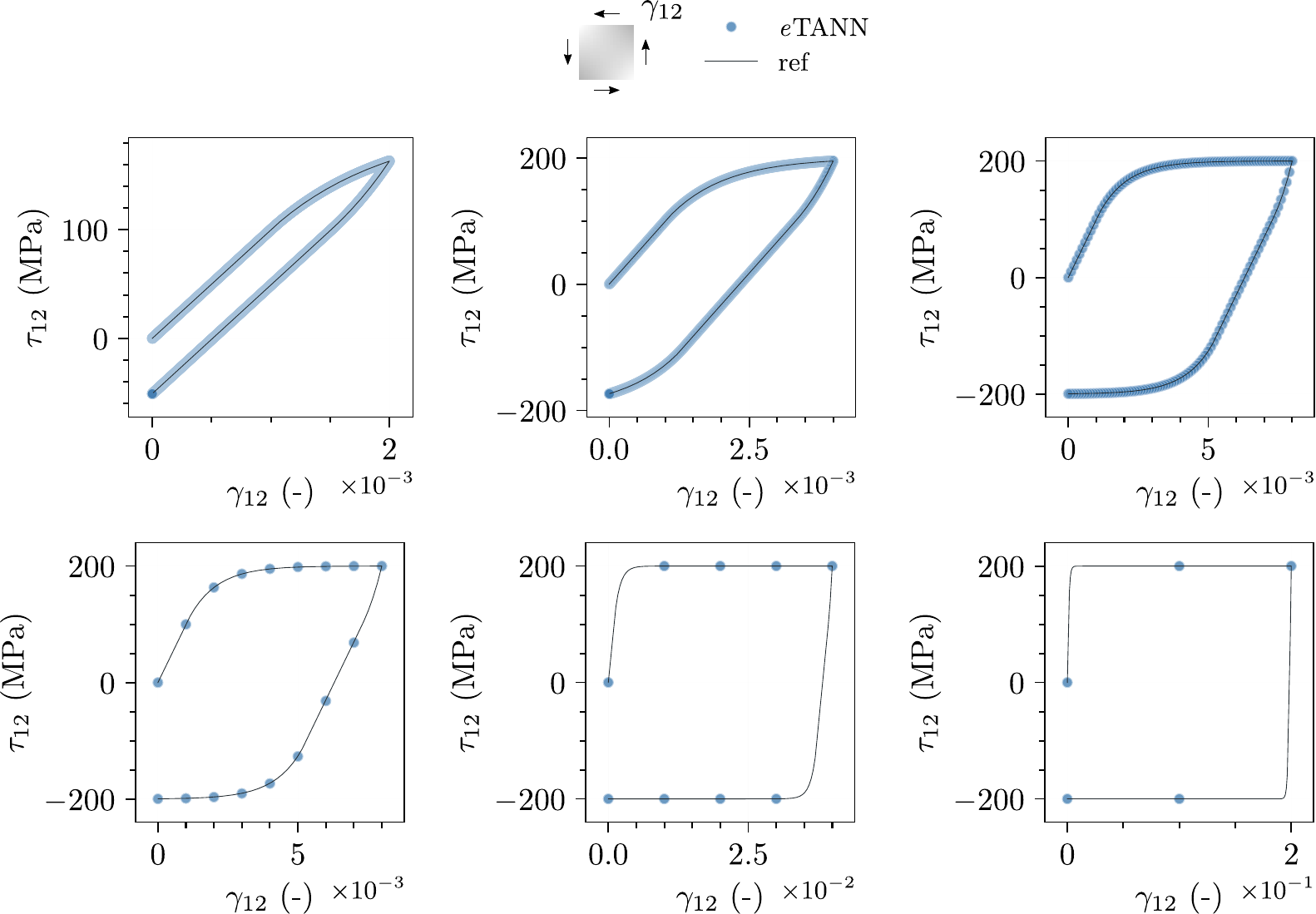}
	\caption{Simple shear loading path at varying of the strain increment ($\Delta \gamma$) and constant shear rate $\dot{\gamma}=100$ s\textsuperscript{-1}. From left to right: $\Delta \gamma=10^{-6},10^{-5},10^{-4}$ (top), $\Delta \gamma=10^{-3},10^{-2},10^{-1}$ (bottom).}
	\label{fig:strain_increment}
\end{figure}

\noindent \hyperref[fig:strain_increment]{Figure~\ref*{fig:strain_increment}} shows the prediction of the network for a pure shear loading, using different strain increments and time steps, demonstrating once more the high generalization capabilities obtained with the proposed approach. Notice how the predictions do not deend on the increment size, contrary to existing approaches \citep[see][among others]{lefik2003artificial,lefik2009artificial}.

\subsubsection{Rate-dependent materials with microstructure}
\label{par:lattice}
\begin{figure}[h]
	\centering
  \includegraphics[width=0.25\linewidth]{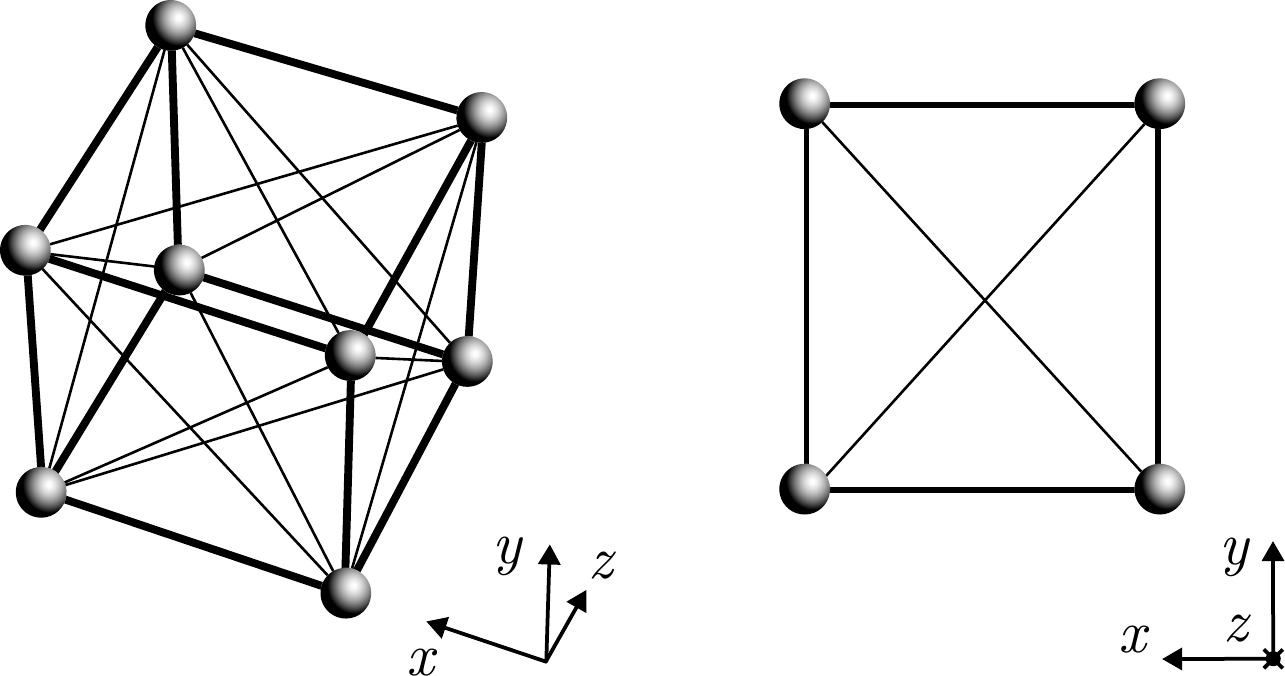}
	\caption{\small Unit lattice cell composed of by 24 bars with elasto-viscoplastic behavior with isotropic hardening.}
	\label{fig:lattice}
\end{figure}

\noindent The last example is a metamaterial with microstructure made of bars with an elasto-viscoplastic behavior and isotropic hardening, see \hyperref[fig:lattice]{Figure~\ref*{fig:lattice}}. The material parameters are kept the same with the first and third example (see \hyperref[par:rateindependent]{Paragraph~\ref*{par:rateindependent}} and \hyperref[par:viscosity]{paragraph~\ref*{par:viscosity}}), except for $\mu=25$ s. The microstructural bars have a constant circular cross-section equal to 1 mm\textsuperscript{2}.\\
Periodic boundary conditions at unit cell are considered in order to use the trained network to solve multiscale simulations in the next paragraph.

The microscopic incremental problem on the lattice cell is solved in terms of the microscopic displacement field, by a code previously developed by the authors \citep{masistefanou2021}. Stress, strain, free energy, and dissipation rate are expressed in terms of volume averages with respect to the unit cell's volume $\mathcal{V}$,
\begin{equation*}
\langle\bm{\sigma} \rangle = \frac{1}{|\mathcal{V}|} \int_{\mathcal{V}} \bm{\sigma} \, dx, \quad 
\langle\bm{\varepsilon} \rangle = \frac{1}{|\mathcal{V}|} \int_{\mathcal{V}} \bm{\varepsilon} \, dx, \quad
\langle\psi \rangle = \frac{1}{|\mathcal{V}|} \int_{\mathcal{V}} \psi \, dx,\quad
\langle d \rangle = \frac{1}{|\mathcal{V}|} \int_{\mathcal{V}} d \, dx.
\end{equation*}
Assuming periodic displacement fields, Eq. (\ref{eq:thermo3}), when expressed in terms of volume averaged quantities, remains of the same form, and so the thermodynamics restrictions, i.e., (\ref{eq:stress}-\ref{eq:dissipation}).

In contrast with the applications in the previous paragraphs, here the internal variables are not a priori determined and the data-driven identification of them is performed used the \textit{e}TANN formulation in \hyperref[fig:archtectureTANN_final]{Figure~\ref*{fig:archtectureTANN_final}}. Following \hyperref[subsec:internalvariableunknown]{Subsection~\ref*{subsec:internalvariableunknown}} \citep[see also][]{masistefanou2021}, a minimum set of internal coordinates is selected. The internal coordinates contain the microscopic total and inelastic fields and the deviatoric invariant of the microscopic inelastic strain, which are enough for describing completely the state of the microstructure.

\subparagraph{Datasets, hyperparameters, and training}
56'000 datasets are generated by applying random strain rate and increments and used to train the network. The 80\textsuperscript{th} and 100\textsuperscript{th} percentiles of the imposed strains and strain rates are $3.7, \times 10^{-4}$ and $3\times 10^{-2}$, and, $1$ and $9$, respectively.

The free energy density network consists of one hidden layer, with 312 nodes, and GELU activation. The evolution equation network has three hidden layers, with 1056 nodes, and GELU activations. The encoder and decoder networks are composed of three hidden layers, with a total of 252 nodes, and hyperbolic tangent activation.

\begin{figure}[h]
	\centering
  \includegraphics[width=0.25\linewidth]{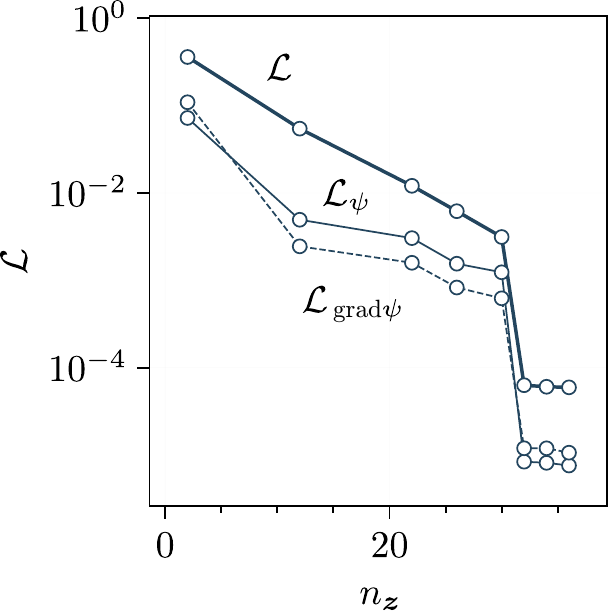}
	\caption{\small Convergence of the training losses at varying of the number of internal variables, $n_{\bm{z}}$.}
	\label{fig:nz}
\end{figure}

In finding an appropriate set of internal variables, the choice of the dimension of the latent representations of the encoder is of paramount importance. Sensitivity analyses with respect to the number of latent representations are needed. In particular, we are interested in identifying the minimum latent dimension such that the training loss $\mathcal{L}$, Eq. (\ref{eq:training_loss}), is minimum. This is shown in \hyperref[fig:nz]{Figure~\ref*{fig:nz}}, where the training losses $\mathcal{L}$, $\mathcal{L}_{\psi}$, and $\mathcal{L}_{\text{grad}\psi}$ are plotted against the number of the internal variables, $n_{\bm{z}}$. We can observe that convergence is reached at $n_{\bm{z}}=33$. Thus, we can confirm that this is the number of internal variables required to adequately describe the material behavior of the lattice cell.

\begin{figure*}[h]
     \centering
     \begin{subfigure}[b]{0.5\textwidth}
         \centering
     	\includegraphics[width=0.9\linewidth]{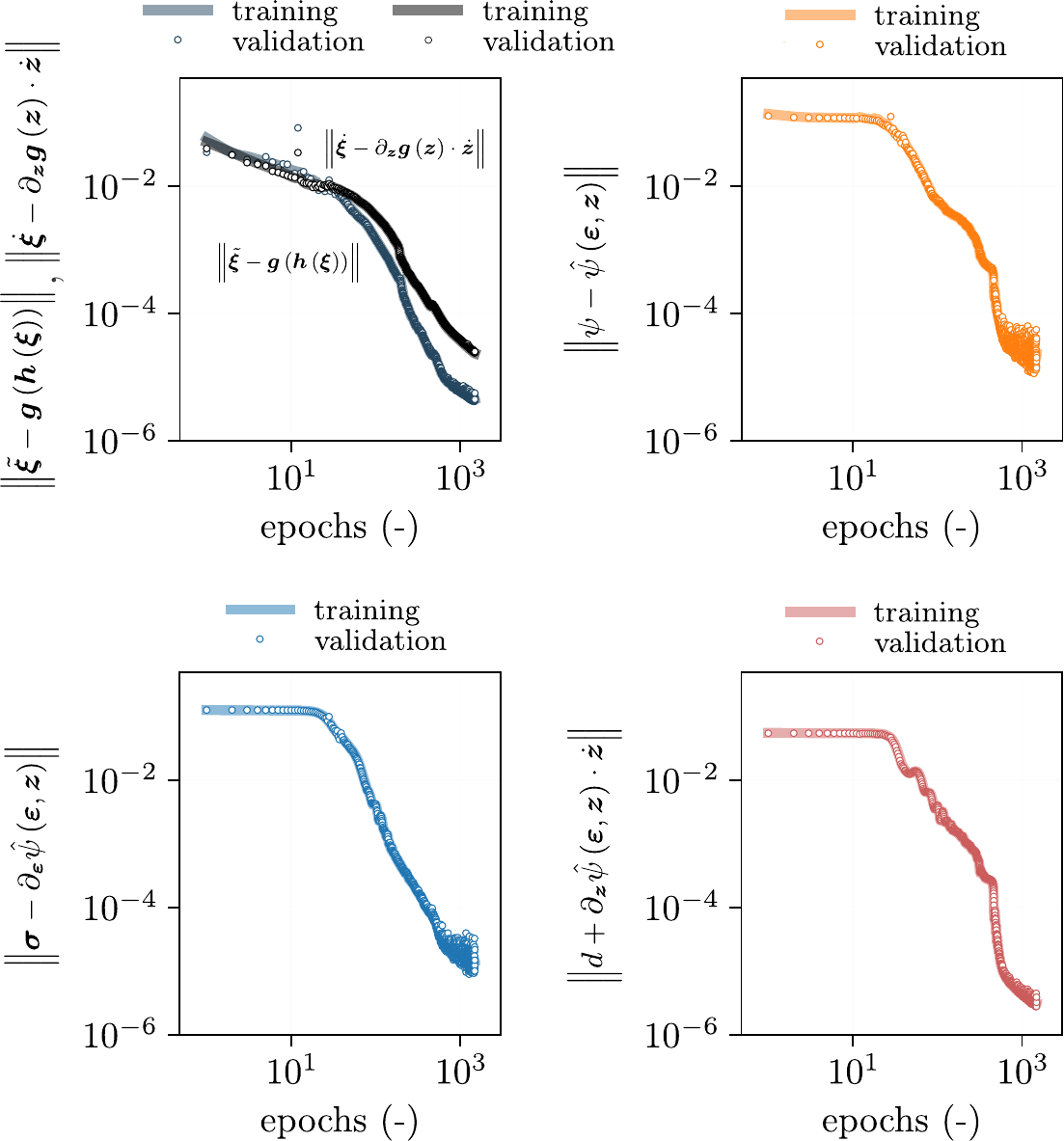}
     	\caption{\scriptsize Training and validation losses of the autoencoder and of the evolution equation of the internal coordinates (left) and of the free energy network (right) during training.}
     \end{subfigure}
     \hspace{0.2cm}
     \begin{subfigure}[b]{0.25\textwidth}
         \centering
     	\includegraphics[width=0.8\linewidth]{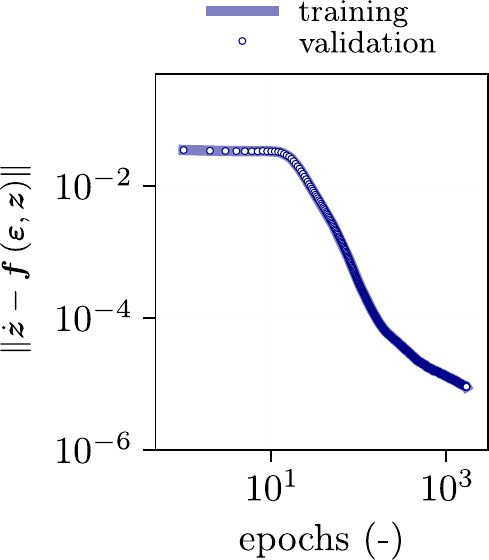}
     	\caption{\scriptsize Training and validation losses of the evolution equation network, during training.}
     \end{subfigure}
        \caption{Training and validation losses for the data-driven identification of internal variables, evolution equations, and material behavior for the lattice cell in \hyperref[fig:lattice]{Figure~\ref*{fig:lattice}}.}
        \label{fig:training_cell}
\end{figure*}

\hyperref[fig:training_cell]{Figure~\ref*{fig:training_cell}} shows the loss, during training, as function of the epochs.

\begin{figure}[h]
  \centering
  \includegraphics[width=0.6\linewidth]{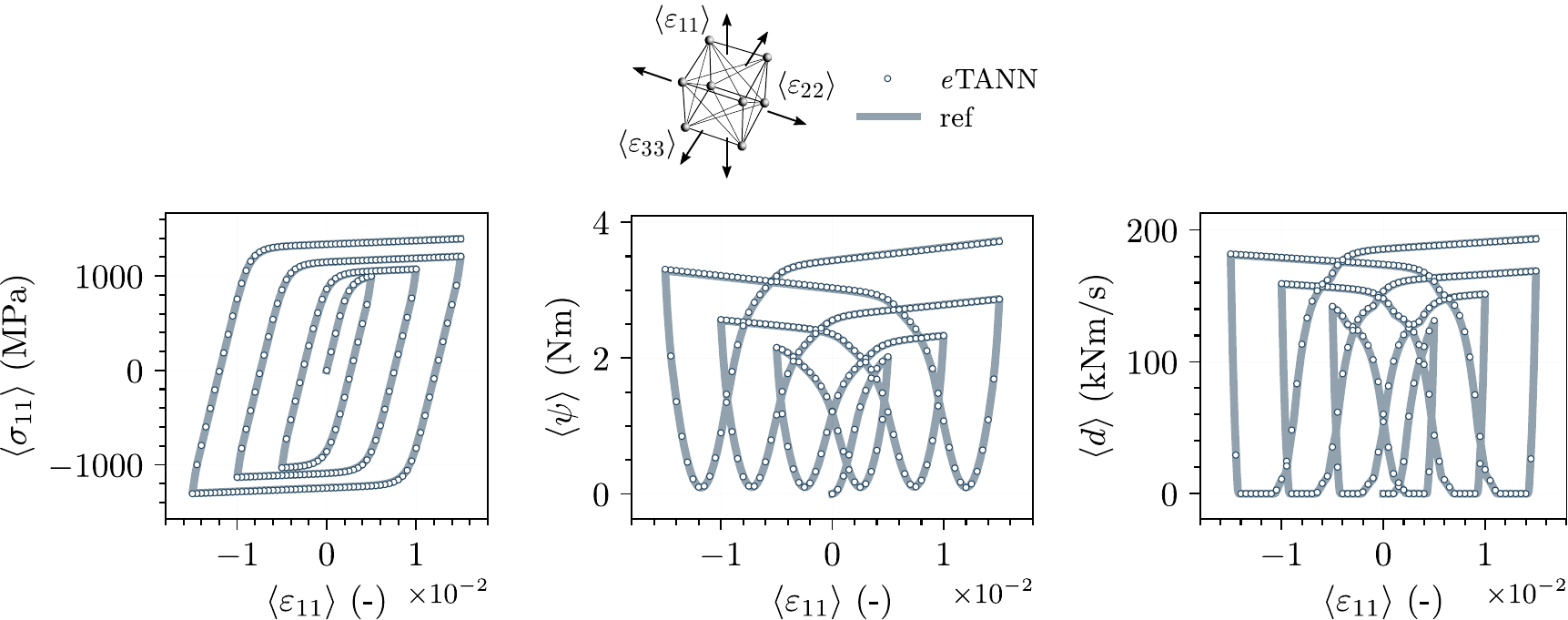}
	\caption{Predictions of the network for a triaxial loading path ($\varepsilon_{22}=\varepsilon_{33}=0.25\varepsilon_{11}$) and strain rate equal to 4 s\textsuperscript{-1}, compared to the reference model. From left to right: stress, free energy, and dissipation rate versus deformation.}
	\label{fig:lattice_biaxial}
\end{figure}

\begin{figure}[h]
  \centering
  \includegraphics[width=0.8\linewidth]{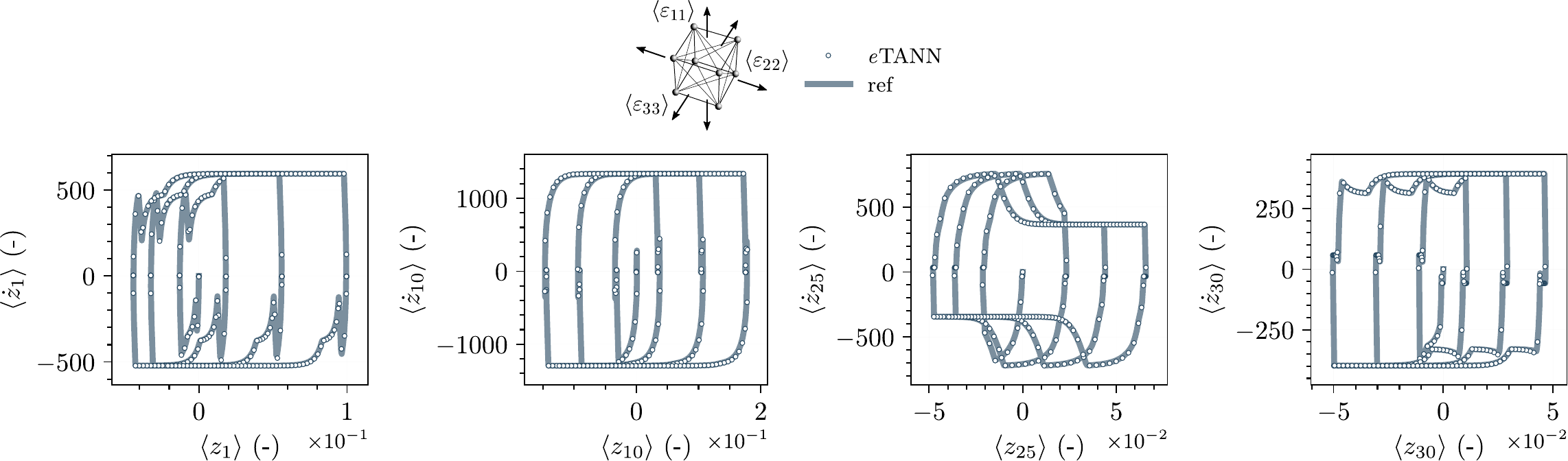}
	\caption{Phase portrait of some of the identified internal variables for a triaxial loading path. The reference curves are obtained by encoding the internal coordinates, whose evolution is computed using the micromechanical model.}
	\label{fig:lattice_biaxial_svars}
\end{figure}

\subparagraph{Results}
We show in \hyperref[fig:lattice_biaxial]{Figure~\ref*{fig:lattice_biaxial}} the predictions the volume average behavior of the metamaterial performed by \textit{e}TANN, with respect to the micromechanical reference model. Despite the demanding triaxial loading path, the model is found to accurately describe the complex material behavior, correctly accounting for both viscosity and hardening. The relative average error in the predicted stress is as low as 0.5\%. In addition, the dissipation rate, computed from the identified internal variables, satisfies the second law of thermodynamics.\\
Particularly interesting is to observe the evolution of the internal variables and their rates, which are shown in \hyperref[fig:lattice_biaxial_svars]{Figure~\ref*{fig:lattice_biaxial_svars}}, for the same triaxial path. The reference values are computed from the encoding of the internal coordinates, as computed by the micromechanical model. The agreement of the predictions demonstrates the ability of the \textit{e}TANN in the constitutive modeling of complex materials.

\begin{figure}[h]
  \centering
  \includegraphics[width=0.6\linewidth]{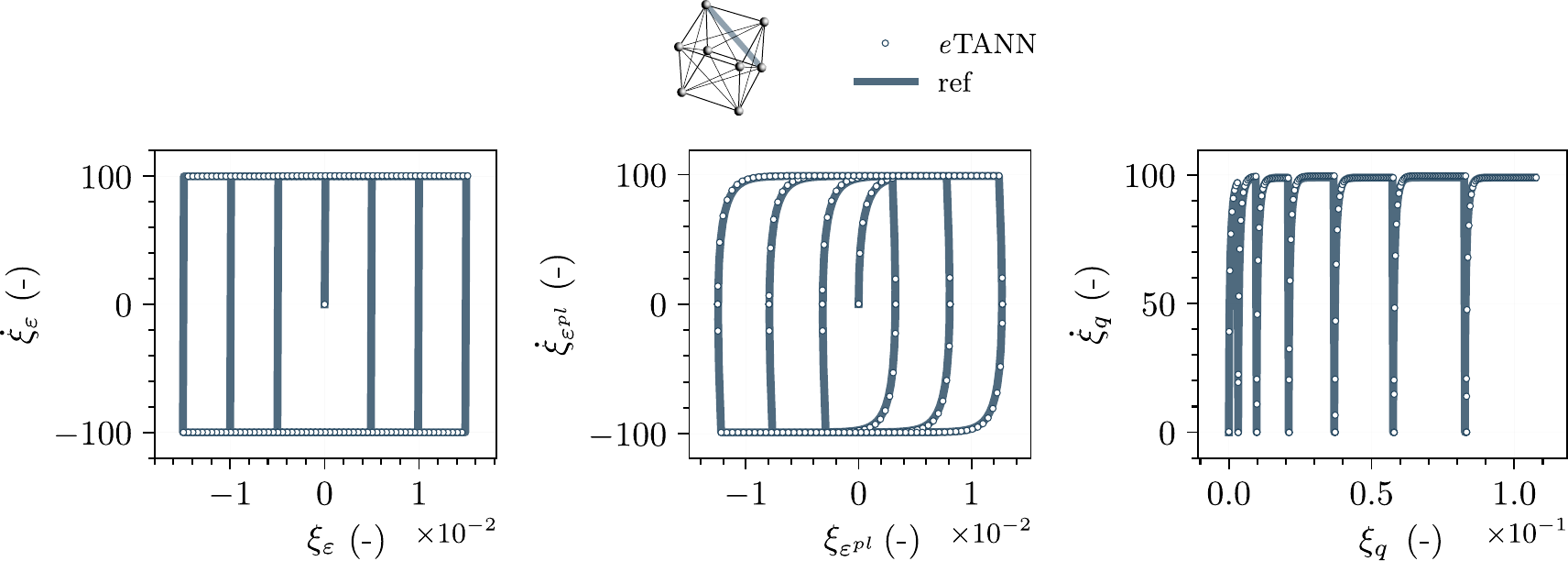}
	\caption{Phase portrait of some of the internal coordinates, related to the highlighted bar in the lattice cell (top). From left to right: micromechanical deformation, micromechanical inelastic deformation, and micromechanical deviatoric part of the accumulated inelastic deformation.}
	\label{fig:lattice_biaxial_ICdot}
\end{figure}

Internal variables and their evolution equation can be decoded to obtain the internal coordinates of the microstructure and their dynamics. This is shown in \hyperref[fig:lattice_biaxial_ICdot]{Figure~\ref*{fig:lattice_biaxial_ICdot}}, for the internal coordinates of one of the microscopic bars (as highlighted in the figure). The approach is found to deliver an accurate description of the microstructure of the material. Note, again, that the internal coordinates and their rates are not computed on-the-fly, at each step, but only after to demonstrate the capability of the proposed approach while maintaining a minimal material description based on the internal variables.  


\begin{figure}[h]
  \centering
  \includegraphics[width=0.8\linewidth]{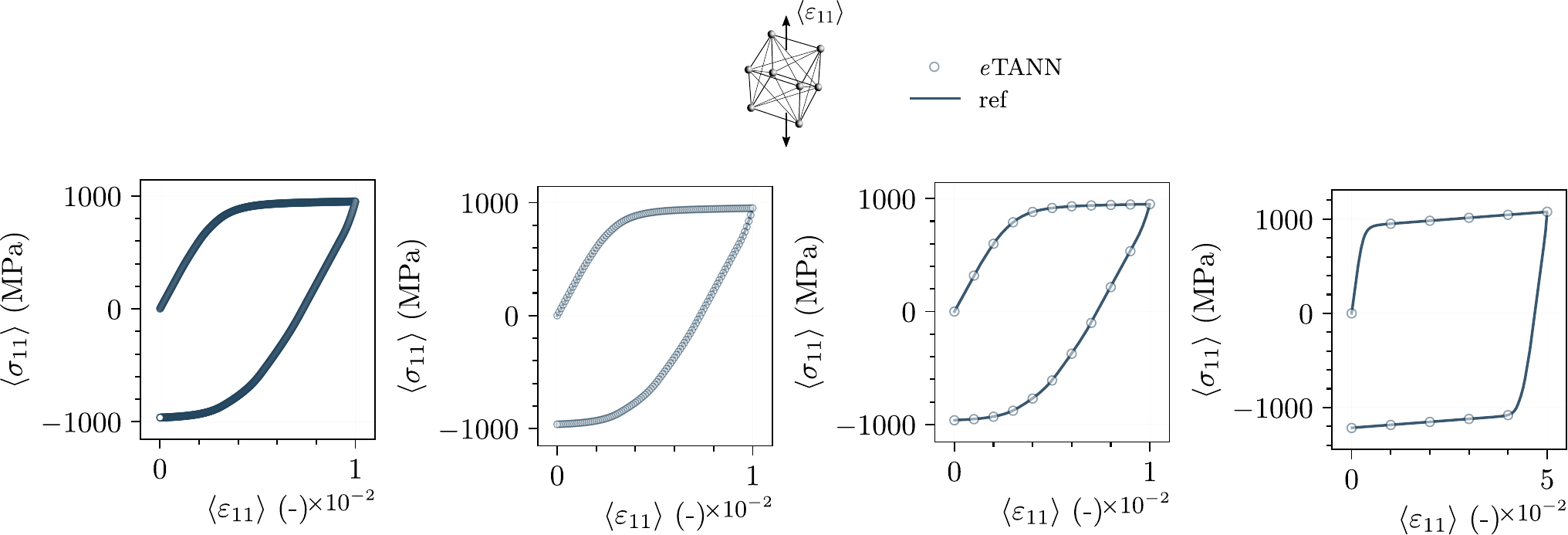}
	\caption{Unixial loading path at varying of the strain increment ($\langle\Delta \varepsilon\rangle$) and constant strain rate $\dot{\varepsilon}=4$ s\textsuperscript{-1}. From left to right: $\langle\Delta \varepsilon\rangle=10^{-5},10^{-4},10^{-3},10^{-2}$.}
	\label{fig:lattice_increments}
\end{figure}

When used to model materials with unknown internal variables, \textit{e}TANN still benefit of the decoupling between the incremental formulation and the material state description. For the sake of completeness, we show in \hyperref[fig:lattice_increments]{Figure~\ref*{fig:lattice_increments}} the prediction of the network for a uniaxial loading, using different strain increments and time steps. Being the predictions independent of such artificial quantities, the proposed method come with high degrees of generalization, independently of the complexity of the material at hand.

\subsection{Applications in computational multiscale modeling}
\label{sec:multiscale}

\noindent The new, proposed approach can be used to predict the behavior of structures composed of complex materials and accelerate state-of-the-art multiscale simulations. Here, we consider a large-scale metastructure, composed of a unit-cell (cf. the unit cell presented in \hyperref[par:lattice]{Paragraph~\ref*{par:lattice}}) periodically distributed in space. For bridging the scales, we use asymptotic homogenization following an incremental formulation \citep{miehe2002strain}. Consequently, we assume the scale-separation hypothesis, i.e., we assume the existence of two independent scales $x$ and $y = x/\epsilon$, with $\epsilon$ being the ratio of the dimension of the unit-cell with respect to the size of the macroscale structure \citep[see][]{sanchez1986homogenization,Bakhvalov1989,pinho2009asymptotic}. 

The application in \hyperref[par:lattice]{Paragraph~\ref*{par:lattice}} considers periodic boundary conditions, thus by construction the network is already trained to predict the solution of the auxiliary problem of homogenization. As a result, \textit{e}TANN can be used, at inference, for performing large, multiscale analyses. This is accomplished relying on the FEM$\times$TANN approach, developed in \citet{masistefanou2021}. In FEM$\times$TANN, we perform Finite Element analyses by a straightforward replacement of classical constitutive models, at the Gauss points, with the trained network (cf. asymptotic homogenization). The tangent matrix is computed, at each Gauss integration point, by virtue of the auto-differentiation of \textit{e}TANN. An extensive study on the achievable computational accelerations obtained with the FEM$\times$TANN approach can be found in \citet{masistefanou2021}. Also note that the approach can benefit of additional computational accelerations adopting adaptive mesh refinements and clustering \citep{zienkiewicz2005finite,stein2004encyclopedia,koutsovasilis2010model,benaimeche2022k}.\\

\begin{figure}[h]
  \centering
  \includegraphics[width=0.8\linewidth]{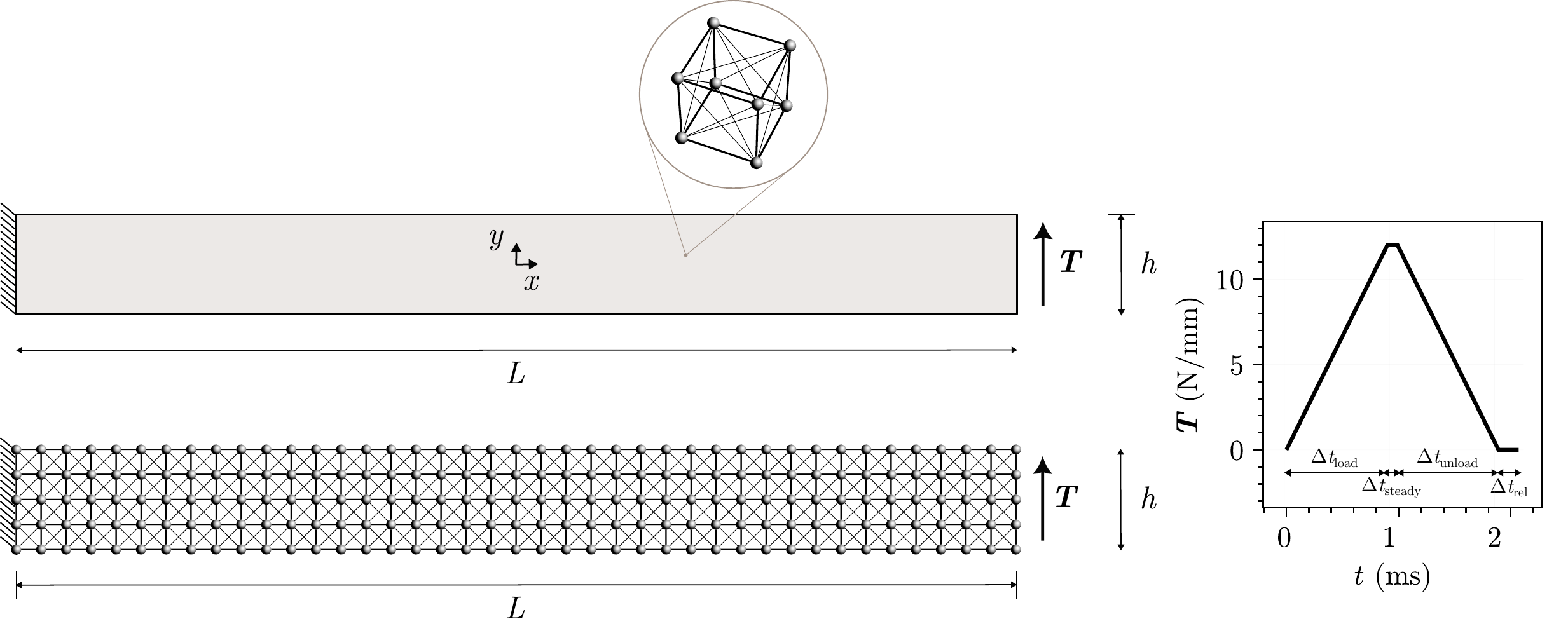}
	\caption{Multiscale and micromechanical model of a panel, with dimensions $L=10$ mm, $h = 2$ mm, under plan strain conditions, subjected to a shearing load.}
	\label{fig:multiscale_model}
\end{figure}

By means of an example, we consider the problem of a panel, clapped on one end, and subjected, quasi-statically, to a shearing load at the other end. \hyperref[fig:multiscale_model]{Figure~\ref*{fig:multiscale_model}} depicts the initial geometry. Plane strain conditions are adopted. Initially, a force equal to 12 N/mm (per unit cell in width) is applied linearly in time until $\Delta t_\text{load}=0.9$ ms. The loading time is selected in order to have strain rates of the order of $1$ s\textsuperscript{$-1$}, corresponding to those developed due to impact loading \citep[see e.g.][]{sadeghi2019scaled}. The force is then maintained over a prescribed time interval $\Delta t_\text{steady} = 0.09$ ms, and, finally, the structure is unloaded gradually within a time interval $\Delta t_\text{unload}=0.9$ ms (see \hyperref[fig:multiscale_model]{Figure~\ref*{fig:multiscale_model}}, right). To capture relaxation effects due to the viscous behavior of the microstructure, the unloaded phase is maintained for an additional time interval $\Delta t_\text{rel}=0.18$ ms. The FE homogenized model consists of 80 linear tetrahedral elements in length and 8 in height (with crossed diagonals). The number of elements was determined by mesh convergence analyses.\\

\begin{figure}[h!]
  \centering
  \begin{subfigure}[b]{0.4\textwidth}
         \centering
     	\includegraphics[width=1\textwidth]{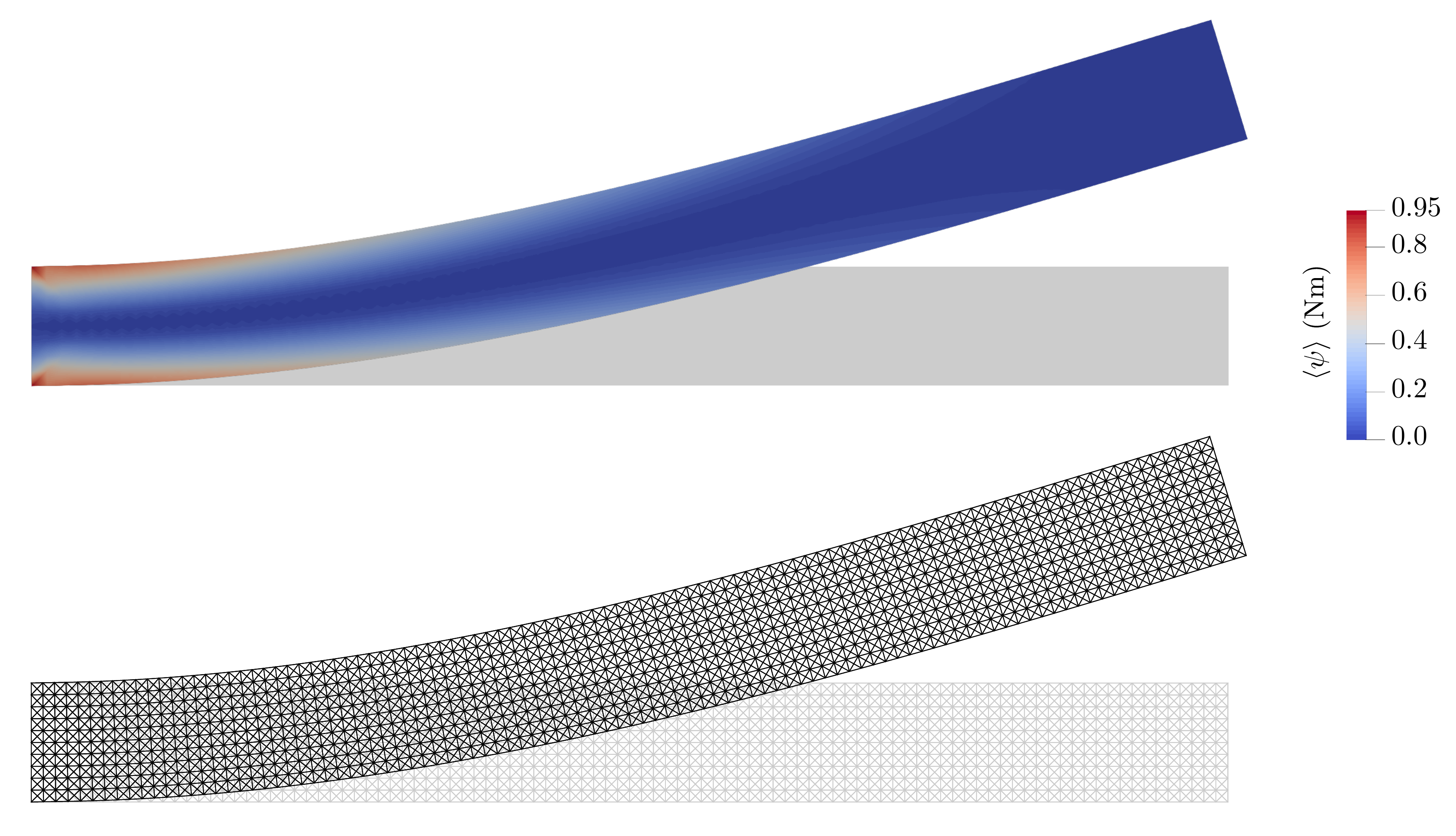}
     	\caption{\scriptsize end of the loading phase: $t=\Delta t_\text{load}$.}
     \end{subfigure}
     \vspace{0.1cm}
     
    \begin{subfigure}[b]{0.4\textwidth}
         \centering
     	\includegraphics[width=1\textwidth]{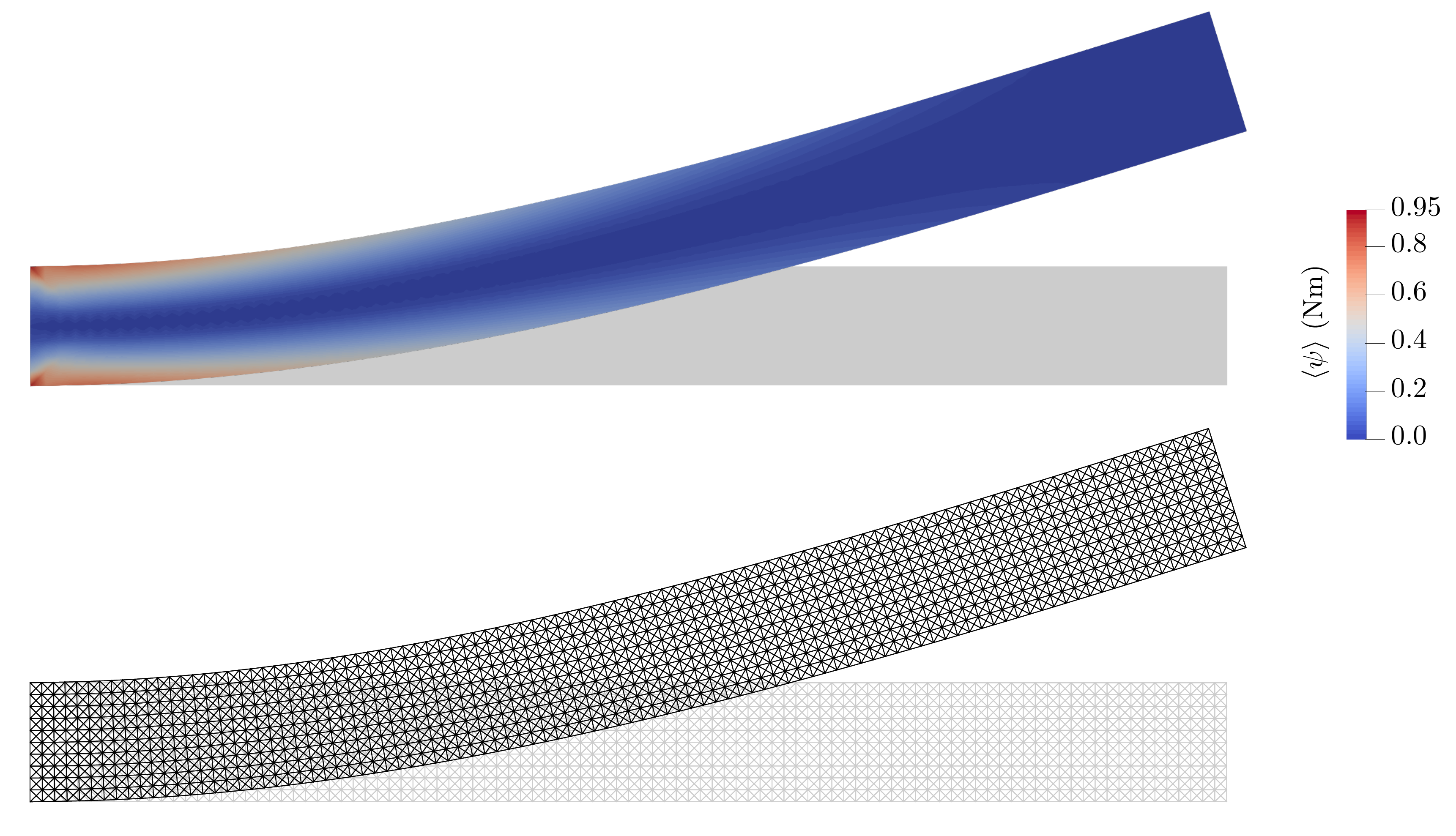}
     	\caption{\scriptsize end of the steady phase: $t=\Delta t_\text{steady}+\Delta t_\text{load}$.}
     \end{subfigure}
     \vspace{0.1cm}
     
     \begin{subfigure}[b]{0.4\textwidth}
         \centering
     	\includegraphics[width=1\textwidth]{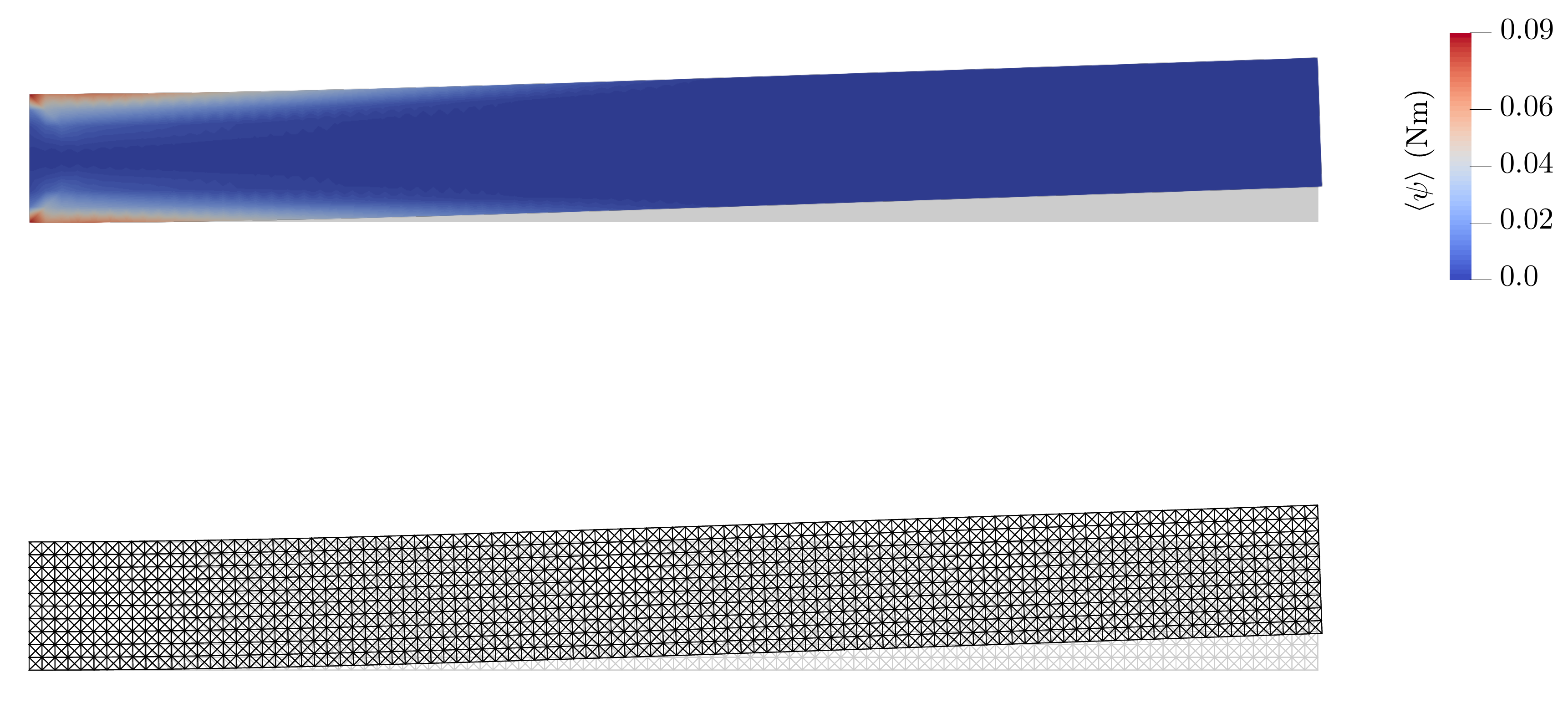}
     	\caption{\scriptsize end of the relaxation phase: $t=\Delta t_\text{steady}+\Delta t_\text{load}+\Delta t_\text{unload} +\Delta t_\text{{rel}}$.}
     \end{subfigure}
     
	\caption{Deformed shapes of the micromechanical ($\epsilon=0.1$) and the homogenized model. Displacement are magnified by factor of 10. Shaded regions represent the initial configuration at $t=0$ ms. The contours show the homogenized elastic energy density, which is locked after unloading due to plastic effects.}
	\label{fig:multiscale_def}
\end{figure}

In the following, we investigate and compare the solutions obtained from the (exact) micromechanical model with those from the homogenized one, with \textit{e}TANN. \\
\hyperref[fig:multiscale_def]{Figure~\ref*{fig:multiscale_def}} depicts the deformed shapes of the micromechanical and homogenized models (magnified by a factor of 10), at the end of the loading (top), steady (middle), and unloading (bottom) phase. Contours of the free energy density in the homogenized model identify the areas where high stresses develop.

\begin{figure}[h]
  \centering
  \begin{subfigure}[b]{0.25\textwidth}
         \centering
     	\includegraphics[width=0.75\textwidth]{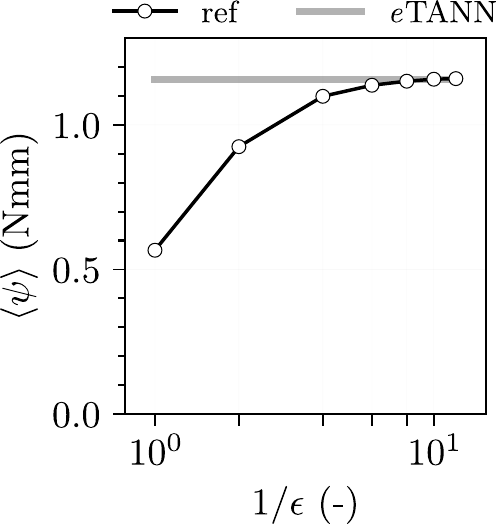}
     	\caption{\scriptsize total energy.}
     \end{subfigure}
     \hspace{0.5cm}     
    \begin{subfigure}[b]{0.25\textwidth}
         \centering
     	\includegraphics[width=0.75\textwidth]{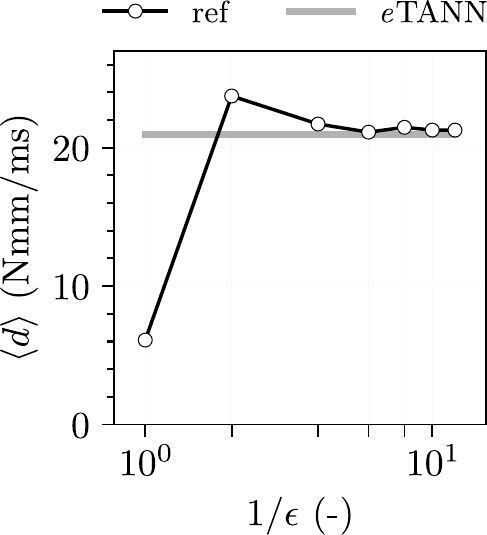}
     	\caption{\scriptsize total dissipation rate.}
     \end{subfigure}
          
	\caption{Total energy and dissipation rate of the micromechanical (ref) and homogenized (\textit{e}TANN) models ($t=\Delta t_\text{load}$), at varying of the unit cell size, $\epsilon$.}
	\label{fig:multiscale_epsilon}
\end{figure}

\noindent By virtue of the asymptotic homogenization \citep{miehe2002strain}, the micromechanical solution should converge to the homogenized one, as $\epsilon$ tends to zero. Accordingly, we compare the results of the micromechanical model, with different sizes of the unit-cell $\epsilon$, with those obtained with the FEM$\times$TANN approach. \hyperref[fig:multiscale_epsilon]{Figure~\ref*{fig:multiscale_epsilon}} depicts the total energy and dissipation rate in function of $\epsilon$.  For $1/\epsilon\geq 6$, the maximum relative error is as low as 1.7\% (in energy) and 2.3\% (in dissipation). The convergence of the response of the homogenized model to the response of the micromechanical simulations is achieved at $\epsilon=0.1$ (with an error approximately equal to 0.5\%), which is taken as the reference solution for the following analyses.\\

\begin{figure}[h]
  \centering
     	\includegraphics[width=0.2\linewidth]{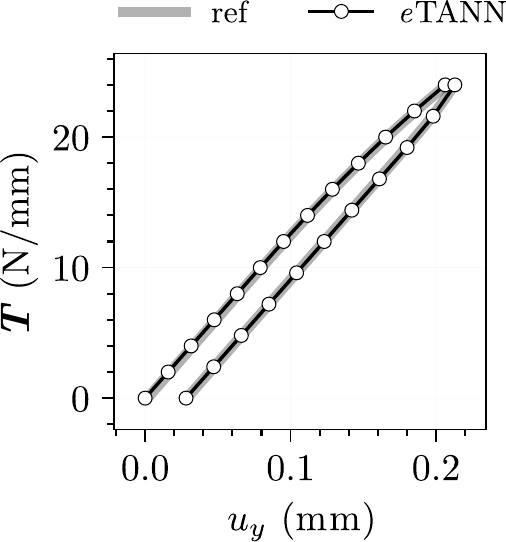}
	\caption{Force-displacement response of the micromechanical ($\epsilon=0.1$) and the homogenized model.}
	\label{fig:multiscale_T}
\end{figure}

As far as it concerns the kinematics of the problem, we compare, in \hyperref[fig:multiscale_T]{Figure~\ref*{fig:multiscale_T}}, the force-displacement response of the homogenized model with the micromechanical reference solution. For the former, we consider the (zeroth-order) approximation of the displacements, while for the latter, the microscopic displacements at the nodes are considered as reference. A good agreement between the two models is observed. At the end of the loading phase, the relative error in the displacements is smaller than 0.4\%, revealing the accuracy of the proposed approach. Reducing further $\epsilon$ would further reduce the error. However the micromechanical simulations take a disproportional  amount of time to complete which renders the comparison hard to be performed.\\
During the steady phase, the material displays viscous effects that alter the displacements and deformation fields. This is captured by the homogenized model with a relative error in the displacement of 0.25\%. In addition, the model correctly predicts the residual (non-zero) vertical displacement (cf. \hyperref[fig:multiscale_T]{Figure~\ref*{fig:multiscale_T}}).\\

\begin{figure}[h]
  \centering
  \begin{subfigure}[b]{0.4\textwidth}
         \centering
     	\includegraphics[width=\linewidth]{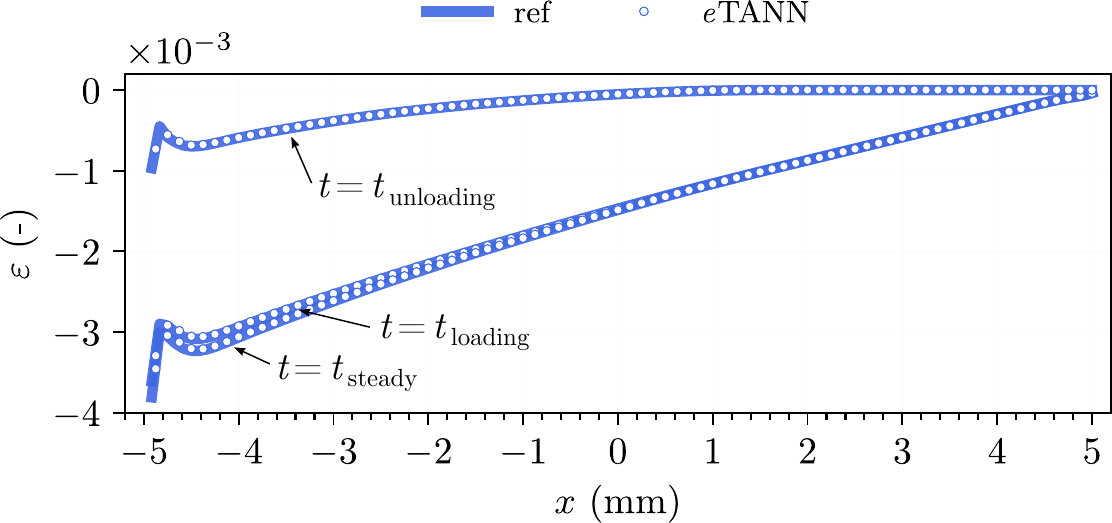}
     	\caption{\scriptsize distribution of the microscopic total deformation.}
     \end{subfigure}
\vspace{0.3cm}
     
     \begin{subfigure}[b]{0.4\textwidth}
         \centering
     	\includegraphics[width=\linewidth]{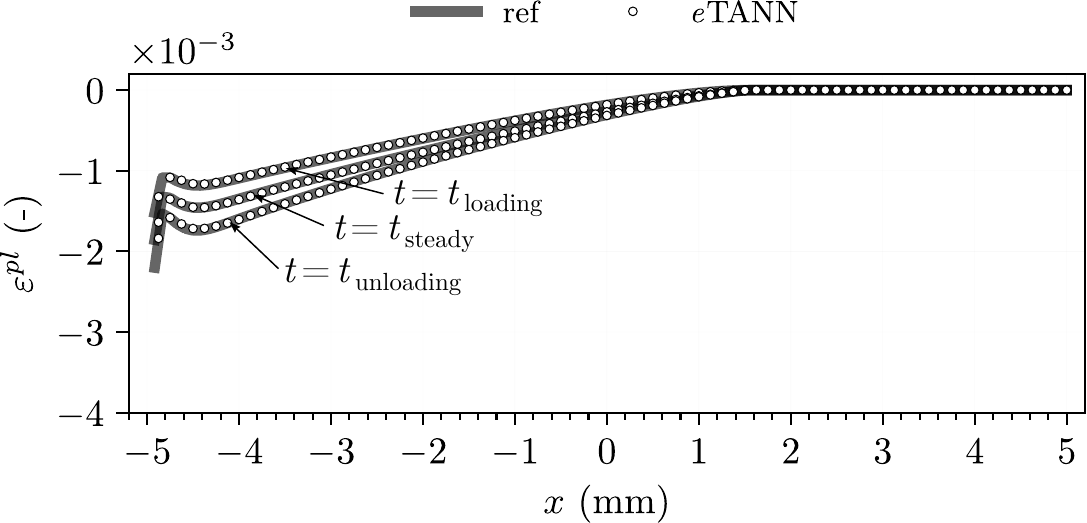}
     	\caption{\scriptsize distribution of the microscopic inelastic deformation.}
     \end{subfigure}
	\caption{Microscopic deformations (total and inelastic) at the intrados of the panel at the end of the loading, steady, and unloading plus relaxation phase. In the micromechanical model ($\epsilon=0.1$), the microscopic deformations are computed from the deformations of the bars with axis parallel to the intrados. In the homogenized model, the microscopic deformations are obtained from the decoding of the identified internal variables at the Gauss points (no interpolation).}
	\label{fig:multiscale_ez}
\end{figure}

Besides the good agreement in terms of average quantities, such as displacements and free energy, \textit{e}TANN allow describing the evolution of microscopic fields, i.e., the internal coordinates $\bm{\xi}$, within the homogenized model. This is what is commonly called \textit{localization} in the frame of asymptotic homogenization. Differently from the classical procedure \citep[see e.g.][]{pinho2009asymptotic}, the microscopic fields within the unit cell can be reconstructed, with the proposed approach, by the straightforward decoding of the identified internal variables--that is, $\tilde{\bm{\xi}}=\bm{g}\left( \bm{z}\right)$. For this purpose, we analyze the distribution of the microscopic deformations at the intrados of the panel. In the micromechanical model ($\epsilon=0.1$), the microscopic deformations are computed from the deformations of those bars whose axis is parallel to the intrados. \hyperref[fig:multiscale_ez]{Figure~\ref*{fig:multiscale_ez}} shows the comparison of the two models in terms of the total and plastic deformations at the microscale at ($i$) the end of the loading phase, ($ii$) end of the steady phase, and ($iii$) the end of the unloading phase.\\
An excellent agreement is found. From one side, the FEM$\times$TANN approach allows to capture with high fidelity the deformation redistribution (increase) during the steady phase due to viscosity. From the other side, the microscopic inelastic deformations show not only an overall agreement with the micromechanical solution but also, and more importantly, an accurate prediction of the location of the plastic region. The transition from the elastic to the plastic region, in the micromechanical model, is located at $x = 1.6$ mm (see \hyperref[fig:multiscale_model]{Figure~\ref*{fig:multiscale_model}}), while the homogenized model predicts it at $x = 1.625$ mm. Note that the homogenized solution is only given at the location of the Gauss integration points, without any interpolation. This may justify the slight difference in the plastic region location.\\
Furthermore, the proposed method allows to capture quite complex micromechanical mechanisms: here, the phenomenon of \textit{trapped} elastic energy (or elastic deformations). Indeed, at the end of the unloading phase, the inelastic deformation is larger than the total microscropic deformation (cf. \hyperref[fig:multiscale_ez]{Figure~\ref*{fig:multiscale_ez}}). This phenomenon takes place due to the fact that elastic deformations remain trapped within the plastified region of the structure, from the clapped end to the end of the plastic region ($x\in [-5,1.625]$ mm) and do not vanish, even if no external force is applied. On the contrary, for $x>1.625$ mm, the total and inelastic deformations are zero, and the elastic deformations vanish at unloading. 
 
It is worth noticing that the predictions at the vicinity of the fixed end ($x=-5$ mm) of the panel display some differences, yet minor ones, with respect to the micromechanical solution. This is due to the presence of high strain gradients whose wavelength has amplitude comparable to the size of the unit-cell. Remedies to alleviate these boundary layer effects exist \citep[see e.g.][]{Bakhvalov1989}, but this is out of scope of the present work. Nevertheless, despite the limit of validity of the asymptotic homogenization theory (at first-order), the FEM$\times$TANN approach with \textit{e}TANN is able to provide excellent results compared to the (exact) micromechanical problem and could be used in higher order homogenization schemes.\\

\begin{figure}[h]
  \centering
  \begin{subfigure}[b]{0.4\textwidth}
         \centering
     	\includegraphics[width=1\textwidth]{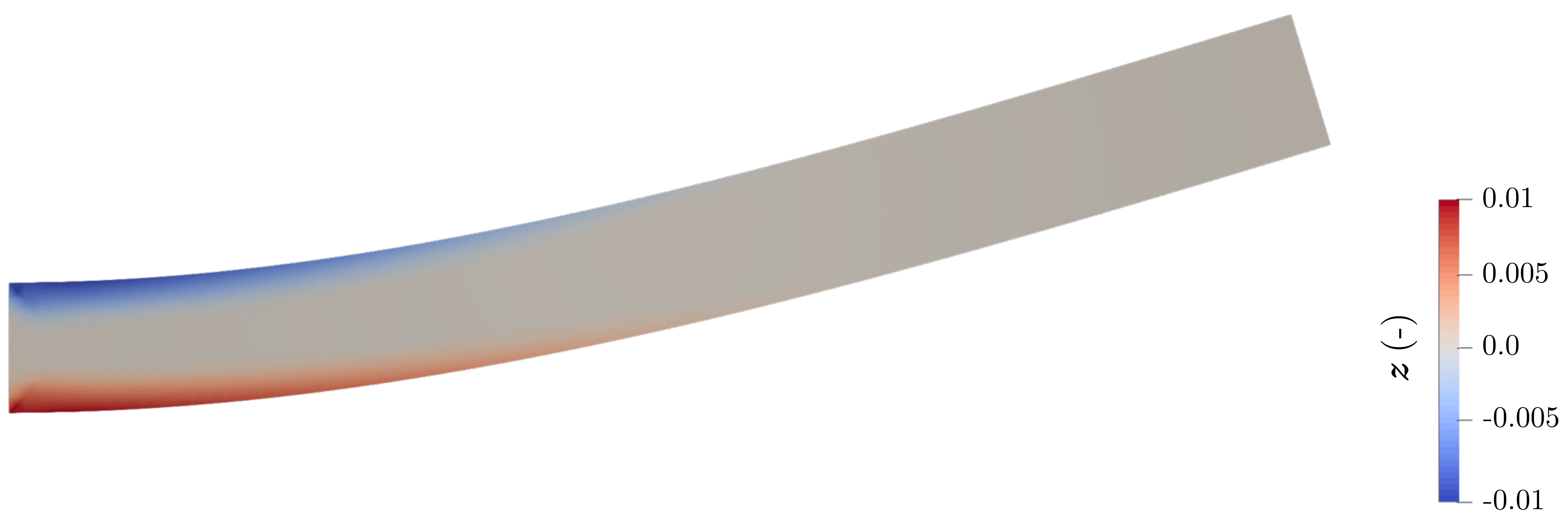}
     	\caption{\scriptsize end of the loading phase.}
     \end{subfigure}     
    \begin{subfigure}[b]{0.4\textwidth}
         \centering
     	\includegraphics[width=1\textwidth]{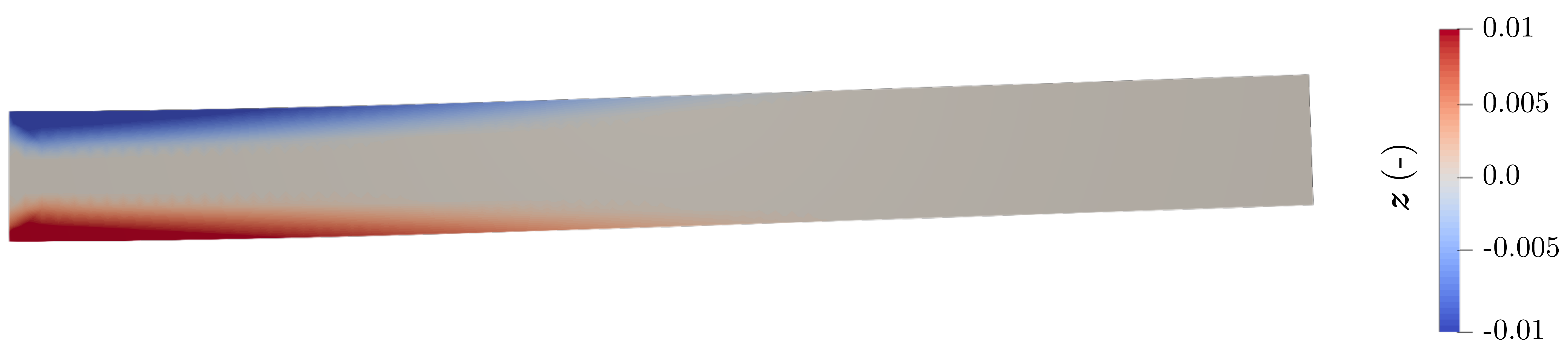}
     	\caption{\scriptsize end of the relaxation phase.}
     \end{subfigure}
          
	\caption{Distribution of one of the identified internal variables at the end of the loading (left) and unloading (right) phase.}
	\label{fig:multiscale_SV}
\end{figure}

For the sake of completeness, \hyperref[fig:multiscale_SV]{Figure~\ref*{fig:multiscale_SV}} depicts the distribution of one of the 33 internal variables. By virtue of the fact that, the residual value, at unloading, is different from zero, we can conclude that the identified internal variables are able to describe the dissipative microscopic phenomena taking place within the metamaterial here considered.

\section{Conclusions}
Deep learning and data-driven approaches have the potential to revolutionize the way we perceive models in mechanics and material science. Particularly inspiring has been the possibility of hardwiring first principles within such approaches and learn constitutive material models that respect the underling physics \citep{karniadakis2021physics,hernandez2021deep,masi2021thermodynamics,klein2022polyconvex,cueto2022thermodynamics}. 

Yet, until now, the generalization capabilities of artificial neural networks have been an obstacle to the development of a rigorous way to describe material behaviors. At the heart of such issue, it is the classical vision in solid mechanics of determining an incremental formulation, i.e. discrete-time, of constitutive models. The translation of this point of view in deep learning has brought neural network structures based not only on physical quantities, but, also on artificial ones, such as strain increments and time steps. This fuzzy mixture and the consequent formulations are major limitations of all existing approaches and prevent the development of scalable and systematic data-driven methods for the modeling of complex materials, displaying path-dependent behavior.

In this work, we proposed a new approach, called evolution Thermodynamics-based Artificial Neural Networks (\textit{e}TANN) which, inspired by the philosophy of TANN \citep{masi2021thermodynamics,masistefanou2021} and the theory of internal variables \citep{coleman1967thermodynamics}, allows the decoupling of the material description from the incremental formulation. \textit{e}TANN give material representations that are continuous-time and, thus, independent of the artificial quantities corresponding to the incremental formulation. This, together with the fact that \textit{e}TANN guarantee, by construction, the fulfillment of the laws of thermodynamics make possible to uncover and describe material behaviors in a scalable manner, offering a high degree of generalization.

The proposed approach finds straightforward implementation to materials for which the internal variables are known. The latter can be selected based on the phenomena and mechanisms which are thought to drive the behavior of a particular material.\\
However, this choice is not always trivial, especially for the case of materials possessing multiple inherent scales, e.g. metamaterials. Accordingly, we developed a strategy which, inspired by previous works \citep{masistefanou2021}, allows to identify, from data and the laws of thermodynamics, admissible set of internal variables, in a general way. The idea is to leverage latent representations of all those quantities describing the microscopic material behavior, called herein internal coordinates, and enforce the fulfillment of first principles. Differently from the developments of TANN in \citet{masistefanou2021}, the new approach allow to identify the evolution equations of both the internal variables and the internal coordinates, rather than the incremental formulation of the latter.\\

Through several applications, we demonstrated the advantages of the formulation local in time of \textit{e}TANN. We showed that our approach can be used to model a broad spectrum of material behaviors, from plasticity to damage and viscosity (and combination of them). In particular, the applications showed that the new formalism allows, independently of the complexity of the material behavior, to be independent of the increment size (in deformations and/or in time), in contrast with all existing approaches. 

Finally, \textit{e}TANN have the potential of replacing the solution of the auxiliary problem in multiscale simulations, by learning both the average and the microscopic behavior of the unit cell. This is demonstrated for the emblematic case of a metamaterial, i.e., a lattice structure. The data-driven identification of the internal variables and evolution equations from the underlying microscopic degrees of freedom allow accurate, scalable, and fast modeling of the behavior of complex materials.\\
By means of an example, we showed that \textit{e}TANN, combined with the mathematical theory of asymptotic homogenization, allow to efficiently and accurately simulate large-scale structures in inelasticity. Besides the enhanced computational accelerations that can be achieved \cite[extensively discussed in][]{masistefanou2021}, here we focused on the high accuracy of the proposed approach in describing both the macroscopic and the microscopic response through detailed comparisons with micromechanical simulations.

Note that the proposed methodology can be also used for micromorphic/generalized continua \citep{germain1973method,forest2020continuum} by simply expanding the stress-strain space. Furthermore, \textit{e}TANN can be used to recover descriptions of any material for which the theory of internal variables hold true. Applications to an even broader range of complex materials is the objective of future works.

\section*{Acknowledgments}
\noindent The authors acknowledge the support of the European Research Council (ERC) under the European Union Horizon 2020 research and innovation program (Grant agreement ID 757848 CoQuake).

\section*{Data availability}
All data is available on request from F.M.

\bibliographystyle{plainnat}
\bibliography{Bibliography}  

\section*{Appendix. Data generation} 
\noindent For each application, material datasets are generated using a stress-point algorithm \citep{ortiz1986analysis,godio2016multisurface}. For all homogeneous materials, except the case with damage, we rely on the libraries implemented in \citet{geolab}. In the case of damage, we use, instead, the material constitutive algorithm implemented in \citet{abaqus,giovanniCFM}. For the metamaterial unit cell, the solution of the auxiliary problem is performed relying on a FE code specifically developed by the authors \citep[see][]{masistefanou2021}.

In all applications, independently of the particular material and behavior, datasets are generated applying random strain increments, within constant time steps. First, we identify an initial reference state for the material at time $t=0$ characterized by zero strains and stresses. Then, random strain increments, considered constant over each time step $\Delta t$, are applied sequentially up to a maximum time $t_f$. In particular, the strain increments are drawn component-wise from random standard distributions with zero mean and standard deviation varying between $5\times10^{-5}$ (damage) and $5\times10^{-4}$ (all other applications). The random strain paths include uniaxial, biaxial, triaxial, and full component loading. At time $t_f$, the initial reference state and the time are re-initialized. This solution is made to avoid excessively large material deformations within the datasets and to fill densely the stress-strain response domain. Finally, the randomly generated datasets are shuffled to diversify data within the batches used for training (see \hyperref[subsec:networks]{Subsection~\ref*{subsec:networks}}).

Once datasets are generated, strain rates are computed as
\begin{equation}
\dot{\bm{\varepsilon}}(t_+)=\dot{\bm{\varepsilon}}\big(\left(t+\Delta t\right)_-\big) = \frac{{\bm{\varepsilon}}\big(\left(t+\Delta t\right)_-\big) -  \bm{\varepsilon}(t_+)}{\Delta t}=\dot{\bm{\varepsilon}}^*,
\end{equation}
by leveraging the fact that strain increments are constant over each $\Delta t$. All other rate quantities are computed relying on a first-order backward finite difference approximation, e.g.
\begin{equation}
\dot{\bm{z}}(t_+) = \frac{{\bm{z}}\big(\left(t+\Delta t\right)_-\big) -  \bm{z}(t_+)}{\Delta t}+ O\left(\Delta t^2\right).
\end{equation}
Note that higher order schemes may be preferred as the proposed approach does not depend on the order of the finite difference approximation.

By leveraging the discretization in time proper to the data generation, the evolution equation network is trained such that
\begin{equation}
\dot{\bm{z}}(t) = \bm{f}\big(\bm{\varepsilon}\left(t\right),\bm{z}\left(t\right) \big) = \bm{f}\big(\bm{\varepsilon}\left(t_+\right)+\dot{\bm{\varepsilon}}^*\left(t-t_+\right),\bm{z}(t) \big),
\end{equation}
where $\dot{\bm{\varepsilon}}^*$ should not misinterpreted as the strain rate $\dot{\bm{\varepsilon}}$, thus the form of $\bm{f}$ in Eq. (\ref{eq:evolution}) remains unchanged, in the sense that does not depend on the strain rate. Therefore, the thermodynamic developments presented in \citep{coleman1967thermodynamics} are respected.

\end{document}